\input{epsf.tex} 
\documentclass[12pt]{article}

\begin{document}
\thispagestyle{empty}
\setcounter{page}{0}
\renewcommand{\theequation}{\thesection.\arabic{equation}} 

{\hfill{ULB-TH/02-21}}

{\hfill{DCPT-02/45}}

{\hfill{\tt hep-th/0207238}}

\vspace{1cm}

\begin{center} {\bf THE BOSONIC MOTHER OF FERMIONIC D-BRANES} 

\vspace{1cm}

Auttakit Chattaraputi${}^a$, Fran\c cois
Englert${}^b$, Laurent
Houart${}^c$\footnote{Research
Associate F.N.R.S.}
\\ and Anne Taormina${}^a$

\footnotesize \vspace{.5 cm}
 ${}^a${\em Department of Mathematical Sciences}, {\em University of
Durham}\\ {\em South Road, DH1 3LE Durham,
England}\\auttakit.chattaraputi@durham.ac.uk\\anne.taormina@durham.ac.uk 

\vspace{.2cm}

${}^b${\em Service de Physique Th\'eorique\\ Universit\'e Libre de Bruxelles,
 Campus Plaine, C.P.225\\Boulevard du Triomphe, B-1050
Bruxelles, Belgium}\\ fenglert@ulb.ac.be

\vspace{.2cm}

${}^c${\em Service de Physique Th\'eorique et Math\'ematique }\\  {\em
Universit\'e Libre de Bruxelles, Campus Plaine C.P. 231}\\ {\em Boulevard du
Triomphe, B-1050 Bruxelles, Belgium}\\ lhouart@ulb.ac.be

\end{center}

\vspace {1cm}
\centerline{ABSTRACT}
\vspace{- 1mm}
\begin{quote}\small  We extend the search  for
fermionic  subspaces of the bosonic string compactified on $E_8 \times
SO(16)$ lattices to include {\em all} fermionic
D-branes.  This extension constraints the truncation procedure previously
proposed  and relates
   the
fermionic strings, supersymmetric or not, to  the global structure of the
$SO(16)$ group. The specific  properties of all the   fermionic
D-branes are found to be encoded in  its universal covering,  whose maximal
toroid defines the configuration space torus of their  mother bosonic  theory.
\end{quote}

\baselineskip18pt
\setcounter{footnote}{0}
\newpage

\setcounter{equation}{0}
\section{Introduction and discussion}

It is well-known that ten-dimensional fermionic strings can be analysed in
terms  of  bosonic operators, a consequence of the boson-fermion equivalence
in two dimensions. The approach taken here is different. We wish to
generalize  previous considerations whereby  the Hilbert space of the
perturbative fermionic  closed strings were obtained as subspaces of the
26-dimensional closed bosonic string theory compactified on suitable
16-dimensional manifolds \cite{cent,ens}. Truncating the bosonic string
Hilbert space to its fermionic subspaces yields the conformal field theories of
these fermionic strings. 

The extension of the truncation procedure to  open bosonic strings  
poses a problem. Constraints  arise
because the holomorphic and antiholomorphic conformal algebras describing
the closed string sectors of the bosonic theory are related at the conformal
boundary accommodating  open strings.  These constraints are easily taken
into account when the torus partition function  is diagonal, and, in reference
\cite{eht},  tadpole-free  
open descendants\footnote{ For a comprehensive review on open strings see
reference 
\cite{sagr}.} of Type $IIB$ and Type $OB$  were obtained  in this way from
 bosonic parents.
 Many properties of these open fermionic strings, such as tadpoles and
anomaly cancellations, Chan-Paton groups, D-brane and orientifold tensions,
were found to be encoded  in the bosonic string and were derived in a simple
way from bosonic considerations alone. 

In order to get {\em all} open sectors of
fermionic strings,  bosonic parents with non-diagonal torus partition functions
must be introduced.  The constraints have then non-trivial consequences. 
We shall find that they  tightly relate all bosonic parents to each other as well
as to their fermionic offsprings.

Let us first recall   how a relation between bosonic and
fermionic closed string theories first suggested by Freund
\cite{freund} was established \cite{cent,ens}, and how it was extended to
open strings for diagonal conformal field theories \cite{eht}.

The basic issue was  to uncover space-time fermions from truncation of a
compactified  bosonic string. This was achieved in the following way:

 \hskip 1cm  1) One compactifies, in the light-cone gauge,  $d=16$ transverse
dimensions on a  torus,  leaving eight non-compact dimensions with 
transverse group\footnote{Throughout the paper, any group locally isomorphic
to the rotational group of dimension $n(n-1)/2$ will be labelled
$SO(n)$. When specifically referring to the universal covering group, we shall
use the notation $\widetilde{SO}(n)$.}
$SO_{trans}(8)$. One chooses a compactification of the
closed string at an enhanced symmetry point with gauge group ${\cal
G}_L\times {\cal G}_R$  where  
${\cal G}_L$ and  ${\cal G}_R$ are semi-simple simply laced Lie groups. Recall
that in terms of the left and right compactified momenta, the mass spectrum
is
\begin{eqnarray} 
\label{spectrum} {\alpha^\prime m_L^2\over 4}&=& \alpha^\prime {\bf p_L}^2
+ N_L -1\, , \nonumber \\ {\alpha^\prime m_R^2\over 4}&=&
\alpha^\prime {\bf p_R}^2 + N_R -1 \, ,
\\ &\hbox{and}&\nonumber\\ m^2 = { m_L^2\over 2}+{ m_R^2\over 2}\quad
&;&\quad m_L^2=m_R^2\, . \end{eqnarray} In Eq.(\ref{spectrum})
$N_L$ and
$N_R$ are the oscillator numbers in 26-dimensions and the zero-modes
$\sqrt{2\alpha^\prime}{\bf p_L}$,
$\sqrt{2\alpha^\prime}{\bf p_R}$, where $\alpha^\prime$ is the string length
squared, span a
$2d$-dimensional even self-dual Lorentzian lattice with negative (resp.
positive) signature for left (resp. right) momenta. This ensures modular
invariance of the closed string spectrum \cite{narain}. For generic toroidal
compactifications, the massless vectors
$\alpha_{-1,R}^\mu ~\alpha_{-1,L}^i\vert 0_L,0_R\rangle$ and
$\alpha_{-1,L}^\mu ~\alpha_{-1,R}^i \vert 0_L,0_R\rangle$, where the indices
$\mu$ and
$i$ respectively refer to non-compact and compact dimensions, generate a
local symmetry $[U_L(1)]^d\times [U_R(1)]^d$. But more massless vectors
arise when $
\sqrt{2\alpha^\prime}{\bf p_L}$ and $\sqrt{2\alpha^\prime}{\bf p_R}$ are roots
of simply laced groups ${\cal G}_L$ and ${\cal G}_R$  of rank $d$ (with root
length
$\sqrt{2}$). The gauge symmetry is enlarged to ${\cal G}_L\times {\cal G}_R$.

For closed strings, the compactification lattice in both sectors  (or in the
right sector only for the heterotic strings) is taken to be a sublattice of the
$E_8\times SO(16) $ weight lattice. This sublattice must preserve modular
invariance, which means that the  left and right compactified momenta
$\sqrt{2\alpha^\prime}{\bf p_L}$,
$\sqrt{2\alpha^\prime}{\bf p_R}$ must span a
$2d$-dimensional even self-dual Lorentzian lattice.

\hskip 1cm  2) This type of compactification produces gauged symmetries in  
subgroups
$SO_{int} (8)$ of $SO(16)$ in both sectors (or in the right sector only for the
heterotic strings). The $SO_{int} (8)$ groups are then mapped onto
$SO_{trans} (8)$ in such a way that the diagonal algebra
$ so_{diag}(8) ={\rm diag} [so_{trans}(8) \times so_{int}(8)]$ becomes
identified with a new transverse algebra. The spinor representations of
$SO_{int} (8)$ describe fermionic states because a rotation in space induces a
half-angle rotation on these states. The consistency of the above
procedure relies on the  possibility of extending the diagonal algebra
$so_{diag}(8)$ to the new full Lorentz algebra
$so_{diag}(9,1)$, a highly non-trivial constraint. To break the original Lorentz
group $SO(25,1)$ in favour of the new one, a truncation consistent with
conformal invariance must be performed on the physical spectrum of the
bosonic string. Actually, the states described by twelve compactified bosonic
fields must be projected out, except for momentum zero-modes of unit length 
\cite{cent,ens}. The removal of twelve bosonic fields accounts for the
difference  between the bosonic and fermionic light cone gauge central
charges. Namely, in units where the central charge of a boson is one, this
difference counts 11 for the superghosts and 
$(1/2) . 2 $ for time-like and longitudinal Majorana fermions. The  zero-modes
of length $\ell=1$ kept in the  twelve truncated dimensions contribute a
constant $\ell^2/2$ to the mass. They account for the removal by truncation of
the oscillator zero-point energies in these  dimensions, namely for
$-(-1/24).12=+1/2 $. 

\hskip 1cm  3) This procedure can be extended to stable open strings
descendants of the Type $IIB$ and Type $OB$ theories  \cite{eht}.  Requiring the
absence of tadpole divergencies in the bosonic string , the
correct Chan-Paton groups for the tadpole-free fermionic strings, including
the anomaly-free $SO(32)$ group of Type I were obtained by truncation. In
addition, the tensions of  space-filling D-branes and of  orientifolds in the
fermionic theories followed from the tensions of their bosonic parents.

To extend the analysis to all fermionic Dp-branes and to open string
descendants of Type $IIA$ and Type $OA$ theories, we  introduce
bosonic parents with non-diagonal  torus partition functions.  All the parent
theories, including the  diagonal theories from which the above results
were obtained, are  related to each other and to the fermionic theories
living in their fermionic subspaces through  the global properties of 
$SO(16)$. Compactification of the bosonic string  on the configuration space
torus defined by the maximal abelian subgroup (maximal toroid) 
$\widetilde{\cal T}$ of  the $ E_8\times \widetilde{SO}(16)$ group yields the 
`mother theory'\footnote{Here and in the rest of the paper the subscript $b$
designates  the bosonic parent theory.} 
$OB_b$
whose fermionic subspace is the Hilbert space of  the non-supersymmetric
$OB$ theory.  Subtori corresponding
 to all distinct  maximal toroids $\widetilde{\cal T}/Z_c$ of the groups locally
isomorphic to $\widetilde {SO}(16)$, where  $Z_c$ labels the distinct
subgroups of the centre of $\widetilde {SO}(16)$,  yield the  parents  of all the 
non-heterotic fermionic strings. The toroid  $\widetilde{\cal T}$ encodes the 
information  about   the specific properties of  {\em all}  stable and unstable
space-filling fermionic D9-branes, {\em all} lower dimensional Dp-branes and
{\em all} tadpole-free descendants.

In Section 2,  all bosonic theories  compactified on $ E_8 \times SO(16)$
lattices, consistent with the boundary constraints,  are obtained. Their
fermionic subspaces in the closed string sector are obtained from a universal
truncation procedure. For $IIB_b$ and $OB_b$  one
recovers the lattice compactifications  previously obtained \cite{eht}.  For
$IIA_b$ and $OA_b$ this is not the case and new lattice compactifications  do
appear.  In Section 3,    the lattice compactifications of the bosonic string
theories $IIB_b, OA_b$ and $IIA_b$ are
 deduced from  the $OB_b$
torus by identifying weight lattice points. All fermionic space-filling
D-branes, stable and unstable, are obtained by truncation of the open
sectors.  Their  characteristic properties,  namely  the partition functions of
open strings ending on a D-brane or joining distinct D-branes, and in particular 
their supersymmetric or non-supersymmetric character,  their charge
conjugation properties and  their tensions, are related  to the global group
properties of the `bosonic' group
$\widetilde {SO}(16)$. In Section 4, we confront the  puzzling problem posed by
the identification of the fermionic subspaces describing even  dimensional
fermionic branes which, in the fermionic theories,  are related to the odd ones
by T-dualities flipping the chirality of space-time fermions. This 
 problem is solved in Section 5. The crucial element underlying the solution is
the existence of `odd' E-dualities which relate bosonic theories differing by
`chiral' elements of $Z_c$. In Section 6, we extend the analysis  of the
tadpole-free descendants given in reference
\cite{eht}  to obtain  the $OA$ orientifold and tadpole-free Chan-Paton group
\cite{sagr}. Mathematical developments are relegated to the appendices.

The identification of all the fermionic subspaces of the bosonic strings given
here, both for closed and open string sectors, establishes a dictionary
translating  conformal field theory properties of all the fermionic strings to
those of  the compactified bosonic string.  We cannot, at this stage, assert that
the very existence of this dictionary  signals a dynamical mechanism whereby 
the fermionic subspaces would separate from the remaining states at the non
perturbative level.  However, as
previously pointed out \cite{cent,ens}, the analogy of the stringy construction
of diagonal subgroups $ so_{diag}(8) $ to obtain space-time fermions out of
bosons with the dynamical formation of fermionic   monopole states in
bosonic field theory
\cite{thooft},  may perhaps be indicative of some similar dynamics. Such 
dynamics would of course be of fundamental importance, as it would
considerably enlarge the scope of the present search for an elusive M-theory.
 Its quest deserves  further investigations.

\setcounter{equation}{0}
\section{The universal symmetric  truncation} 

In this section we construct the fermionic subspaces of the compactified 
closed bosonic string.  We focus our attention on a given sector, say the  right
one.

Consider the closed bosonic string compactified on a 16-dimensional
torus at an enhanced symmetry point with gauge group ${\cal G}_L\times {\cal
G}_R$ and ${\cal G}_R= E_8
\times SO(16)$. 

We recall the structure of the weight lattice of the $SO(16)$ and $E_8$ groups. 
Generically, the weight lattice of $SO(4m)$ is partitioned into  four cosets 
with respect to the
$2m$-dimensional    root lattice.  These are the conjugacy classes
$(o)_{4m}\, ,(v)_{4m}\, , (s)_{4m}\, ,(c)_{4m}$ which,  under addition, 
 are isomorphic to the centre $Z_2\times Z_2$ of the universal covering 
group  $\widetilde{SO}(4m)$.  The
$(o)_{4m}$ lattice is the root lattice generated by the simple roots.  It
contains the neutral element
$(0,0, ...,0)$. The
$(v)_{4m}$ lattice is the vector lattice whose  smallest weights are $4m$
vectors of norm one; in an orthonormal basis, these are
$\sqrt{2\alpha^\prime}{\bf p_v}=(\pm1,0,...,0)$ and permutations. The
$(s)_{4m}$ and
$(c)_{4m}$ lattices are spinor lattices whose smallest weights have  norm
squared $m/2$ and are the    $2^{2m}$ vectors
$\sqrt{2\alpha^\prime}{\bf p_{s,c}}=(\pm1/2,\pm1/2,...,\pm1/2)$ with even (for
class $(s)_{4m}$ ) or odd (for class $ (c)_{4m}$ ) number of minus signs.  The
$E_8$ lattice is self-dual  and its root lattice $(o)_{E_8}$ may be expressed as
$(o)_{16} + (s)_{16}$.

The  closed string partition function is modular invariant. Its 
${\cal G}_L\times {\cal G}_R$ lattice contribution $P(\tau,\bar\tau)$ is
separately modular invariant and given by

\begin{equation}
\label{factor} P(\tau,\bar \tau)=\sum_{\alpha,\beta} N_{\alpha\beta} \
\bar
\alpha_L(\bar\tau)\ \beta_R(\tau)\, ,
\end{equation} where 
\begin{equation}
\label{partition}
\beta_R (\tau)=\sum_{\sqrt{2\alpha^\prime}{\bf p}_{oR}\in \, (o)} \exp
\{2\pi i\tau[ \alpha^\prime({\bf p}_{o R}+{\bf p}_{\beta R} )^2 +
N_R^{(c)} -{\delta
\over 24 }]\}\, .
\end{equation} Here $\beta$ is a partition function for a sublattice
$(\beta)$ of the ${\cal G}_R= E_8 \times SO(16)$ weight lattice (i.e.
$(\beta)=(o)_{E_8} \oplus (i)_{16}, \: i=o,v,s,c$) and
${\bf p}_{\beta R}$ is a fixed vector, arbitrarily chosen, of the sublattice
$(\beta)$.
$ N_R^{(c)}$ is the oscillator number in the $\delta=16$ compact dimensions. A
similar expression holds for $\bar
\alpha_L(\bar\tau)$,  $\bar\alpha$ labeling a partition function for a sublattice
of the weight lattice of ${\cal G}_L$. The coefficients
$N_{\alpha\beta}$ are 0 or 1  and are chosen in such a way that 
$P(\tau,\bar \tau)$ is modular invariant.

We now perform the truncation in the right sector. We decompose the
$SO(16)$ factor of ${\cal G}_R$ in $SO^{\, \prime}(8)
\times SO(8)$ and first truncate all states created by oscillators in the 12
dimensions defined by the $E_8\times SO^{\,\prime}(8)$ root lattice.  We then
identify the group
$SO(8)$ with the internal symmetry group $SO(8)_{int}$ defined in the 
introduction. As pointed out there, the closure of the new Lorentz algebra
dictates we keep zero-modes in the 16 compact dimensions in such a way that
\begin{equation}
\label{ghost} \alpha^\prime{\bf p_R}^2[E_8 \times SO(16)] =
\alpha^\prime{\bf p_R}^2[SO(8)] +{1\over 2}\, .
\end{equation} The zero-mode contribution $\frac{1}{2}$ in Eq.(\ref{ghost})
comes from
$SO^{\, \prime}(8)$ as there are no vectors of norm squared one in $E_8$.

The decomposition of an
$SO(16)$ lattice in terms of $SO^{\,\prime}(8) \times SO(8)$ lattices yields
\begin{eqnarray} (o)_{16} = [(o)_{8^\prime}\oplus (o)_{8}] &+&
[(v)_{8^\prime}\oplus (v)_{8}] \, ,\nonumber\\ (v)_{16} = [(v)_{8^\prime}\oplus
(o)_{8}] &+& [(o)_{8^\prime}\oplus (v)_{8}]\nonumber\, ,\\ (s)_{16} =
[(s)_{8^\prime}\oplus (s)_{8}]& +&[ (c)_{8^\prime}\oplus
(c)_{8}]\nonumber\, ,\\ (c)_{16} = [(s)_{8^\prime}\oplus (c)_{8}] &+&
[(c)_{8^\prime}\oplus (s)_{8}]\, .
\end{eqnarray} The vectors of norm one in $SO^{\, \prime}(8)$ are the
4-vectors
$\sqrt{2\alpha^\prime}{\bf p^\prime_v}, \sqrt{2\alpha^\prime}{\bf
p^\prime_s}$ and
$\sqrt{2\alpha^\prime}{\bf p^\prime_c}$ as described above. We choose one
vector
$\sqrt{2\alpha^\prime}{\bf p^\prime_v}$ and one vector
$\sqrt{2\alpha^\prime}{\bf p^\prime_s}$. One may equivalently choose
$\sqrt{2\alpha^\prime}{\bf p^\prime_c}$ instead of
$\sqrt{2\alpha^\prime}{\bf p^\prime_s}$.

Focusing on the first choice, we get the following truncation for the lattice
partition functions
\begin{eqnarray}
\label{truncations} &&o_{16} \rightarrow v_8\, , \qquad v_{16}\rightarrow o_8
\,
,\nonumber\\ &&s_{16} \rightarrow -s_8\, ,
\quad ~ c_{16}
\rightarrow -c_8 \, .
\end{eqnarray} It follows from the closure of the Lorentz algebra that states
belonging to 
$v_8$ or $o_8$ are bosons while those belonging to the spinor partition
functions $s_8$ and
$c_8$ are space-time fermions. In accordance with the spin-statistic theorem
we  have  flipped the sign in the partition function of the space-time spinor
partition functions\footnote{Note that this sign flip is consistent with the
positive sign for the bosonic partition functions because of a sign
ambiguity for the factors in  the decomposition of  the $SO(16)$ lattice
partition function in sum of products of 
 $SO^{\, \prime}(8)
$ and  $SO(8)$  partition functions.}. 

The truncation Eq.(\ref{truncations}) does preserve the modular invariance of 
the original  bosonic closed string theory.  Under the $S$ modular
transformation
$\tau\to -1/\tau$ the  partition functions of the four
 sublattices $(o)_{16},(v)_{16},(s)_{16},(c)_{16}$ transform as
\begin{eqnarray}
\label{sixteen} o_{16} &\rightarrow& {1\over2} [\,
o_{16}+v_{16}+s_{16}+c_{16}\,]
\, ,\nonumber\\ v_{16} &\rightarrow& {1\over2} [\,
o_{16}+v_{16}-s_{16}-c_{16}\,]
\, ,\nonumber\\ s_{16} &\rightarrow& {1\over2} [\,
o_{16}-v_{16}+s_{16}-c_{16}\,]
\, ,\nonumber\\ c_{16} &\rightarrow& {1\over2} [\,
o_{16}-v_{16}-s_{16}+c_{16}\,] \, ,
\end{eqnarray}   as do the four $SO(8)$ partition functions
$o_8,v_8,s_8$ and
$c_8$. From Eq.(\ref{sixteen}) and  the truncation-flip Eq.(\ref{truncations}), it
is indeed easily proven \cite{eht,ns} that  modular invariance is preserved
under truncation, namely that the truncation-flip commutes with both
$S(\tau\to -1/\tau)$ and  $T (\tau\to \tau+1)$.

 The modular invariant compactifications with gauge symmetry  ${\cal
G}_L\times {\cal G}_R$ and ${\cal G}_R=  E_8 \times SO(16)$ yield all
heterotic strings, supersymmetric or not,  when the right sector is  truncated
according to Eq.(\ref{truncations}) \cite{lls,eht}. The other fermionic closed
strings theories, namely Type
$IIB, IIA, OB$ and $OA$, can be obtained from compactifications with 
${\cal G}\times {\cal G}$ symmetry.  Such compactifications are modular
invariant
 for any semi-simple simply laced group ${\cal G}$ of rank
$d$ , if both
$\sqrt{2\alpha^\prime}{\bf p_L}$ and $\sqrt{2\alpha^\prime}{\bf p_R}$ span the
full weight lattice
$\Lambda_{weight}$ of
${\cal G}$, but are constrained to be in the same conjugacy class \cite{en}. Such
lattices will be referred to as EN lattices. Taking  the EN lattice of
${\cal G}=E_8\times SO (16)$ and using Eq.(\ref{truncations}) in both sectors,
one gets the non-supersymmetric OB theory. The same procedure applied to
the EN lattice of ${\cal G}=E_8\times E_8$ yields the supersymmetric $IIB$
theory. The $OA$ and $IIA$ theories follow from the same bosonic theory by
using Eq.(\ref{truncations}) in the right sector and interchanging $s_8$ with 
$c_8$ in the left one. This amounts to use $\sqrt{2\alpha^\prime}{\bf
p^\prime_c}$ instead of
$\sqrt{2\alpha^\prime}{\bf p^\prime_s}$ as the momentum kept in the left
$SO^{\, \prime}(8)$ group.  The subsequent analysis will however show that
while the above asymmetric  truncation for type $IIA$ and $OA$ is valid as long
as only closed strings are considered, it cannot be used for analyzing their
D-branes. We shall then see how to formulate a symmetric truncation for all
cases. This will allow for a  generalization of the D-brane analysis given in
reference
\cite{eht}. More importantly  perhaps, it will reveal relations between 
fermionic theories encoded in their bosonic parents.

The four lattice partition functions $o_{16},v_{16}, s_{16}, c_{16}$  are the
characters of level one representations of the affine Kac-Moody algebra
$\widehat {so}(16)$.  They define an  extended conformal algebra and all
chiral operators describing  interactions of the bosonic string in the right
sector must obey the corresponding fusion rules. These are
 determined by the group multiplication table  of the
conjugacy classes of $SO(16)$, that is of the centre $Z_2\times Z_2$ of its
universal covering. They are given
by
\begin{eqnarray}
\label{bfusion} &\ &
[o_{16}][o_{16}]=[v_{16}][v_{16}]=[s_{16}][s_{16}]=[c_{16}][c_{16}] = [o_{16}]\, ,
\nonumber \\ &\ & [o_{16}][v_{16}]=[v_{16}],\ [o_{16}][s_{16}]=[s_{16}], \
[o_{16}][c_{16}] =[c_{16}]\, ,
\nonumber \\ &\ & [v_{16}][s_{16}]=[c_{16}], \ [v_{16}][c_{16}]=[s_{16}],\
[s_{16}][c_{16}] = [v_{16}]\, .
\end{eqnarray}
In this sector, the chiral operators describing   the
ten-dimensional fermionic string interactions can, in the light-cone gauge,
be expressed in terms of bosonic operators acting on  the
fermionic subspace. These operators must therefore obey the fusion
rules obtained by truncating the bosonic ones,
\begin{eqnarray}
\label{ffusion} &\ & [v_{8}][v_{8}]=[o_{8}][o_{8}]=[s_{8}][s_{8}]=[c_{8}][c_{8}] =
[v_{8}]\, ,
\nonumber \\ &\ & [v_{8}][o_{8}]=[o_{8}],\ [v_{8}][s_{8}]=[s_{8}], \ [v_{8}][c_{8}]
= [c_{8}]\, ,
\nonumber \\ &\ & [o_{8}][s_{8}]=[c_{8}], \ [o_{8}][c_{8}]=[s_{8}],\ [s_{8}][c_{8}]
= [o_{8}]\, .
\end{eqnarray}
In the fusion rules Eq.(\ref{ffusion}) the r\^ole of the identity 
has been transferred from
$[o_{16}]$ to    the vector  class
$[v_8]$    \cite{ens,revlsw}.

Similarly,  for the left extended conformal algebra the fusion rules are 
\begin{eqnarray}
\label{bantifusion} &\ & [\bar o_{16}][\bar o_{16}]=[\bar v_{16}][\bar
v_{16}]=[\bar s_{16}][\bar s_{16}]=[\bar c_{16}][\bar c_{16}] = [\bar o_{16}]\, ,
\nonumber \\ &\ & [\bar o_{16}][\bar v_{16}]=[\bar v_{16}]\, ,\ [\bar
o_{16}][\bar s_{16}]=[\bar s_{16}]\, ,
\ [\bar o_{16}][\bar c_{16}] =[\bar c_{16}]\, ,
\nonumber \\ &\ & [\bar v_{16}][\bar s_{16}]=[\bar c_{16}]\, , \ [\bar
v_{16}][\bar c_{16}]=[\bar s_{16}]\, ,\ [\bar s_{16}][\bar c_{16}] = [\bar v_{16}]\,
.
\end{eqnarray} Under both the symmetric and asymmetric truncations,  the
fermionic fusion rules in the left sector are  the antiholomorphic counterpart
of the fusion rules Eq.(\ref{ffusion}).  Thus, as far as fusion rules are concerned
symmetric and  asymmetric  truncations appear to be valid.

When D-branes are introduced the asymmetric truncation runs into problems.
Indeed in that case, boundary conditions relate the left and  right conformal
families of the bosonic theories leaving only conformal families of a single
algebra. The asymmetric  truncations yielding type $IIA$ and $OA$  are then at
best ambiguous and in  fact turn out to be inconsistent for Dirichlet boundary
conditions.   To obtain all fermionic D-branes   we
must use the {\em symmetric} truncation and thus take Eq.(\ref{truncations}) 
in both sectors of the closed bosonic strings.

To realize this program we cannot restrict ourselves to EN lattices, and we
must find new modular invariant compactifications on left and right  $E_8
\times SO(16)$ lattices. Here and in the rest of the  paper, we only
write down the $SO(16)$ characters in the integrand of  amplitudes:
the `dummy' $E_8$  character and the contribution of the eight non-compact
dimensions are not displayed.   There are {\it four} such modular invariant
partition functions, given by
\begin{eqnarray}
\label{BOB} OB_b&=&\bar o_{16}\ o_{16}+ \bar v_{16}\ v_{16}+
\bar s_{16}\ s_{16}+\bar c_{16}\ c_{16}\, ,\\ \label{BOA} OA_b&=&\bar o_{16}\
o_{16}+ \bar v_{16}\ v_{16}+
\bar s_{16}\ c_{16}+\bar c_{16}\ s_{16}\, ,\\
\label{B2B} IIB_b&=&\bar o_{16}\ o_{16}+ \bar s_{16}\ o_{16}+
\bar o_{16}\ s_{16}+\bar s_{16}\ s_{16}\, ,\\ \label{B2A} IIA_b&=&\bar o_{16}\
o_{16}+ \bar c_{16}\ o_{16}+
\bar o_{16}\ s_{16}+\bar c_{16}\ s_{16}\, .
\end{eqnarray} The $OB_b$ and the $IIB_b$ theories describe the
previously mentioned compactifications on  the EN lattice of
$SO(16)$ and on its $E_8$ sublattice.  One can show by inspection  that the  two
other theories, namely
$OA_b$ and
$IIA_b$, correspond to even self-dual Lorentzian lattices and are thus
modular invariant too. This property also follows from the existence of 
conformal $\sigma$-model actions. Namely the four compactifications
Eqs.(\ref {BOB})-(\ref {B2A}) may be expressed in terms of  actions
\cite{nsw} (from now on we choose
$\alpha^\prime =1/2$)

\begin{eqnarray} S={-1\over 2
\pi}\int d\sigma d\tau \left[\{g_{ab}\partial_\alpha X^a\partial^\alpha X^b +
b_{ab} \epsilon^{\alpha \beta} \partial_\alpha X^a\partial_\beta X^b\}\right.
\nonumber\\
\left. +\eta_{\mu\nu}\partial_\alpha X^\mu\partial^\alpha X^\nu \right]\, ,
\label{action}
\end{eqnarray} with $g_{ab}$ a constant metric and $b_{ab} $ a constant
antisymmetric tensor in the compact directions ($a,b=1,...,16$),
$\eta_{\mu\nu} = (-1;+1,...)$ for
$\mu, \nu =1,...,10$ and
$0\leq \sigma \leq \pi$. The fields $X^a$ are periodic  with period $2\pi$. In this
formalism the left and right momenta are given by 
\begin{eqnarray} {\bf p_R} &=& [{1\over2} m_b + n^a (b_{ab} + g_{ab})] {\bf
e^b},
\ \
\nonumber \\ {\bf p_L} &=&  [{1\over2} m_b + n^a (b_{ab} - g_{ab})] {\bf e^b},
\label{close}
\end{eqnarray} where $\{{\bf e^a}\}$ is the lattice-dual basis of the basis $\{{\bf
e_a}\}$ defining the configuration space torus
\begin{equation}
\label{ptorus} {\bf x} \equiv {\bf x} +2 \pi n^a {\bf e_a} \qquad n^a \in {\cal Z}\ , 
\end{equation} and the   lattice metric is given by \begin{equation}
\label{pmetric} g_{ab}= {\bf e_a. e_b}\ .
\end{equation} 
  Explicit forms of the
$g_{ab}$ and $b_{ab}$ tensors for the four models will be given later.

Using the {\em same} universal truncation
Eq.(\ref{truncations}), we now truncate these theories  in both left and right
sectors . We get 
\begin{eqnarray}
\label{OB} OB_b \rightarrow \bar o_8\ o_8+ \bar v_8\ v_8+
\bar s_8\ s_8+\bar c_8\ c_8 &\equiv& OB\, ,\\ \label{OA} OA_b
\rightarrow
\bar o_8\ o_8+ \bar v_8\ v_8+
\bar s_8\ c_8+\bar c_8\ s_8 &\equiv& OA\, ,\\ \label{2B} IIB_b \rightarrow
\bar v_8\ v_8 - \bar s_8\ v_8 -
\bar v_8\ s_8+\bar s_8\ s_8 &\equiv& IIB\, ,\\
\label{2A} IIA_b \rightarrow \bar v_8\ v_8 - \bar c_8\ v_8 -
\bar v_8\ s_8+\bar c_8\ s_8 &\equiv& IIA\, . \end{eqnarray} Starting from the
four possible bosonic string theories compactified on left and right
$E_8
\times SO(16)$ lattices we thus obtain the four consistent non-heterotic
10-dimensional fermionic theories.  The compactification Eq.(\ref{BOB})
on the  ${\cal G}=E_8\times SO(16)$   EN lattice, which defines the  $OB_b$
theory, yields the same gauge group as the $OA_b$ theory defined by the
compactification Eq.(\ref{BOA}), namely ${\cal G}\times {\cal G}$ with ${\cal
G}=E_8\times SO(16)$. Similarly the compactification Eq.(\ref{B2B}) on the
${\cal G}=E_8\times E_8$ EN lattice  has the same gauge group as the $IIA_b$ 
theory defined by the compactification Eq.(\ref{B2A}), namely  ${\cal G}\times
{\cal G}$ with ${\cal G}=E_8\times E_8$. Comparaison of Eqs.(\ref{OB}) and
(\ref{OA}) with Eqs.(\ref{2B}) and (\ref{2A}) shows that
the enhanced gauge symmetry in the bosonic parents from $SO(16)$ to $E_8$ 
signals the onset of supersymmetry for their fermionic offsprings.  This
correspondence between supersymmetry  in fermionic strings  and the
occurrence of an $E_8$ gauge symmetry  for their bosonic parents (modulo the
`dummy' $E_8$) holds for all closed string theories, including the heterotic
ones \cite{eht}. In the next sections it will be seen that the `breaking' of
supersymmetry is {\em always}, both in closed and
open string sectors, translated to the  breaking of the
$E_8$ symmetry to its subgroup $SO(16)$ in the bosonic parent theories.  This
feature  is indicative of the fundamental r\^ole played in  the bosonic theory
by the
$SO(16)$  weight lattice, of which the $E_8$ lattice is a sublattice, in linking
together their fermionic subspaces.

\setcounter{equation}{0}
\section{D9-branes and  torus geometry} 

In this section and in the following  ones we  discuss the  bosonic 
 D-branes  of the four different bosonic theories compactified on  $E_8 \times
SO(16)$  lattices.  D-branes  break the ${\cal G}\times {\cal G}$
 symmetry of the closed strings to  ${\cal G}$ and may even break ${\cal
G}$ itself. We shall indeed
 find  that, when the unbroken ${\cal G}$ is $E_8$ (modulo the `dummy' 
$E_8$),  breaking of $E_8$ to $SO(16)$ may occur leading to a concomitant
breaking of supersymmetry in the fermionic sector. Such breaking  will
manifest itself by the appearance of the partition function $v_{16}$ in
the open string sector of the $IIB_b$ (or the $IIA_b$) theory (see Table IV
below), and hence, from the truncation $v_{16}\to o_8$, by a tachyon   in the
$IIB$ (or the $IIA$) theory. 

In this section, we relate the properties of the bosonic
D9-branes to the geometry of the configuration space torus characterizing
each compactification. These tori are  linked to each other through global
properties of the universal covering group
$\widetilde{SO}(16)$. We then perform the symmetric truncation and obtain
the space-filling D9-branes of the different fermionic theories.

We want to find  the  bosonic D9-branes with Dirichlet boundary conditions in
the 16-dimensional compact space.   The amplitudes ${\cal A}_{tree}$  
describing  the D9-branes in the  tree channel are obtained from the torus
partition functions Eqs.(\ref{BOB})-(\ref{B2A}) by imposing  Dirichlet boundary
conditions on the compact space.  For open strings the latter   
 do not depend on
$b_{ab}$ and are given by
 \begin{equation}
\label{diro}
\partial_\tau X^a=0\, ,
\end{equation} where $\tau$ is the worldsheet time coordinate and
$\sigma$ the space one.  Using the worldsheet duality which interchanges the
r\^oles of
$\tau$ and
$\sigma$, these equations yield the following relation between the left  and
right momenta:
\begin{equation}
\label{dirc} {\bf p_L}- {\bf p_R}=0\, ,
\end{equation} as well as a match between left and right oscillators in the tree
channel.
 The conditions Eq.(\ref{dirc}) determine the closed strings which propagate in
the annulus amplitude. Imposing them on the four tori  amounts to keep {\em
all} characters which appear diagonally in   Eqs.(\ref{BOB})-(\ref{B2A}).  Up to a
normalization $\alpha$, the annulus amplitudes, written as  closed string tree
amplitudes, are
\begin{eqnarray}
\label{ATB} {\cal A}_{tree}(OB_b)&=& \alpha_{OB_b} (o_{16}+ v_{16}+ s_{16}+
c_{16})\, ,\nonumber\\ {\cal A}_{tree} (OA_b)&=&
\alpha_{OA_b}(o_{16}+
 v_{16})\, ,\nonumber\\ {\cal A}_{tree}(IIB_b)&=&\alpha_{IIB_b}(o_{16}+
s_{16})\, ,\nonumber\\ {\cal A}_{tree}(IIA_b)&=&\alpha_{IIA_b}
~o_{16}\, . \end{eqnarray}

The normalization is determined by rewriting the amplitudes Eq.(\ref{ATB}) as
loop  amplitudes of open strings and by expressing  that these partition
functions count the states of a single string (i.e. without Chan-Paton
multiplicity).  To express ${\cal A}_{tree}$ as a loop amplitude
${\cal A}$, one performs a change of variable and the S-transformation on the
modular parameter ($\tau \rightarrow -1/\tau$). We get\footnote {The factor
$2^5$ comes from the change of variable.} 
\begin{eqnarray}
\label{AB} {\cal A}(OB_b)&=& 2^5\, \alpha_{OB_b} (2\, o_{16} )\, ,\nonumber\\
{\cal A}(OA_b)&=& 2^5\, \alpha_{OA_b} (o_{16}+ v_{16})\, ,\nonumber\\ {\cal
A}(IIB_b)&=&  2^5\, \alpha_{IIB_b} ( o_{16}+ s_{16})\, ,\nonumber\\ {\cal
A}(IIA_b)&=& 2^5\,
\alpha_{IIA_b}(1/2)\, (o_{16}+v_{16}+s_{16}+c_{16})\, .
\end{eqnarray}  To describe, in each theory, {\it one}  elementary D9-brane, we
see that we have to choose $\alpha_{OB_b}=2^{-6},  \alpha_{OA_b}
=\alpha_{IIB_b} =2^{-5}$ and $\alpha_{IIA_b}=2^{-4}$.  Table I gives the
amplitudes of the corresponding elementary
 D9-branes.

\vskip .5cm
\begin{center}
\begin{tabular}{||c||c|c||}
\hline &$2^5{\cal A}_{tree}$&${\cal A}$\\
\hline\hline
$OB_b$&$(1/2)\, (o_{16}+ v_{16}+ s_{16}+ c_{16})$&$o_{16}$\\
\hline
$OA_b$& $o_{16}+ v_{16}$&$o_{16}+ v_{16}$\\
\hline
$IIB_b$& $o_{16}+ s_{16}$&$o_{16}+ s_{16}$\\
\hline
$IIA_b$& $2\, o_{16}$&$o_{16}+ v_{16}+ s_{16}+ c_{16}$\\
\hline
\end{tabular}
\vskip .5cm
 Table I
\end{center}

We now use these results to determine the configuration space torus 
 on which each of the four bosonic theories Eqs.(\ref{BOB})-(\ref{B2A}) is
defined.  The tori define  lattices with basis   vectors
$\{ 2\pi {\bf e_a}\}$ according to Eq.(\ref{ptorus}). In order to find the
lattice corresponding to each theory, we first note that the Dirichlet
condition Eq.(\ref{dirc}) reduces Eq.(\ref{close}) to ${\bf p_L} ={\bf p_R} =
(1/2) m_a {\bf e^a}$ (independent of $b_{ab}$). Using the general expression
for lattice partition functions Eq.(\ref{partition}), we read off for each
model the dual of its $SO(16)$ weight
sublattice from the four tree amplitudes in  Table I.   We
then  deduce the $\{{\bf e_a}\}$ from the duality between the root lattice 
$(o)_{16}$ and the weight lattice
$(o)_{16}+(v)_{16}+(c)_{16}+(s)_{16}$, and from the self-duality of
$(o)_{16}+ (v)_{16}$ and $(o)_{16}+(s)_{16}$. We get
\begin{equation}
\label{pbase} {\bf e_a}= (1/2){\bf w_a}\, ,
\end{equation} where the ${\bf w_a}$ are weight vectors forming a basis of a
sublattice $(r)_{16}$ of the weight lattice of $SO(16)$. The sublattice
$(r)_{16}$ for each theory is \begin{eqnarray}
\label{sublattice} (OB_b)& :& (r)_{16}=(o)_{16}\, , \nonumber\\ (OA_b)& :& 
(r)_{16}=(o)_{16}+(v)_{16}\, ,\nonumber
\\ (IIB_b)& : &(r)_{16}=(o)_{16}+(s)_{16}\, ,\nonumber \\ (IIA_b)& :&
 (r)_{16}=(o)_{16}+(v)_{16}+(s)_{16}+(c)_{16}\, .
\end{eqnarray}
Alternatively we may derive these results from the  loop  channel and
read off the ${\bf e_a}$ directly from Table I.  Indeed,  for Dirichlet   boundary
conditions,  the Hamiltonian eigenvalues for open strings  are independent of
$b_{ab}$ and, for zero-modes,  are equal to
$(2 {\bf l}. 2 {\bf l})/2$ where
${\bf l}=n^a {\bf e_a}$ is the winding lattice.  Comparing the amplitudes  ${\cal
A}$ in Table I to the general expression for lattice partition functions
Eq.(\ref{partition}), we see that
$2 {\bf e_a}={\bf w_a}\, ,$ in accordance with Eq.(\ref{pbase}).

Many properties of these tori can be visualized by a suitable projection.
\vskip.3cm
\hskip 3cm\epsfbox{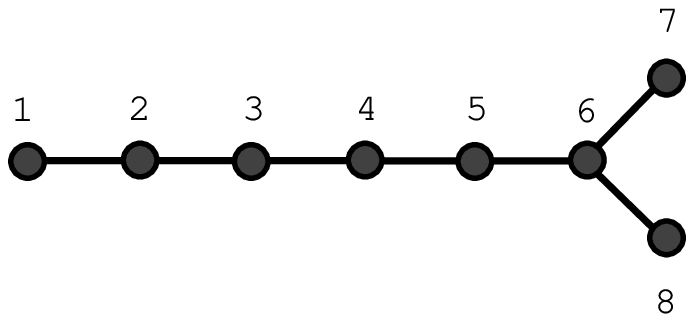}
\begin{quote}\begin{center}
\baselineskip 14pt {\small Fig.1.  Dynkin diagram of $SO(16)$.}\end{center}
\end{quote}
\vskip .3cm
Consider the $SO(16)$ Dynkin diagram in Fig.1.  We project the $SO(16)$ weight
lattice on the plane of the simple roots ${\bf r}_7=(0,0,0,0,0,0,1,-1)$ and
${\bf r}_8=(0,0,0,0,0,0,1,1)$,  along the simple roots ${\bf r}_1, ... ,{\bf r}_6$.
This projection maps the $SO(16)$ onto the   $SO(4)$ weight lattice: $(o_{16}\to
o_4, v_{16}\to v_4, s_{16}\to c_4, c_{16}\to s_4)$.  This is easily verified by
computing the projections ${\bf w}^i_{\, p}$  of the fundamental  weights  ${\bf
w}^i\, (i=1, ... ,7,8)\, $:
 ${\bf w}^i_{\, p}=({\bf w}^i.{\bf w}^7)\,  {\bf r}_7+({\bf w}^i.{\bf w}^8)\, {\bf r}_8
$.

\hskip -.2cm\epsfbox{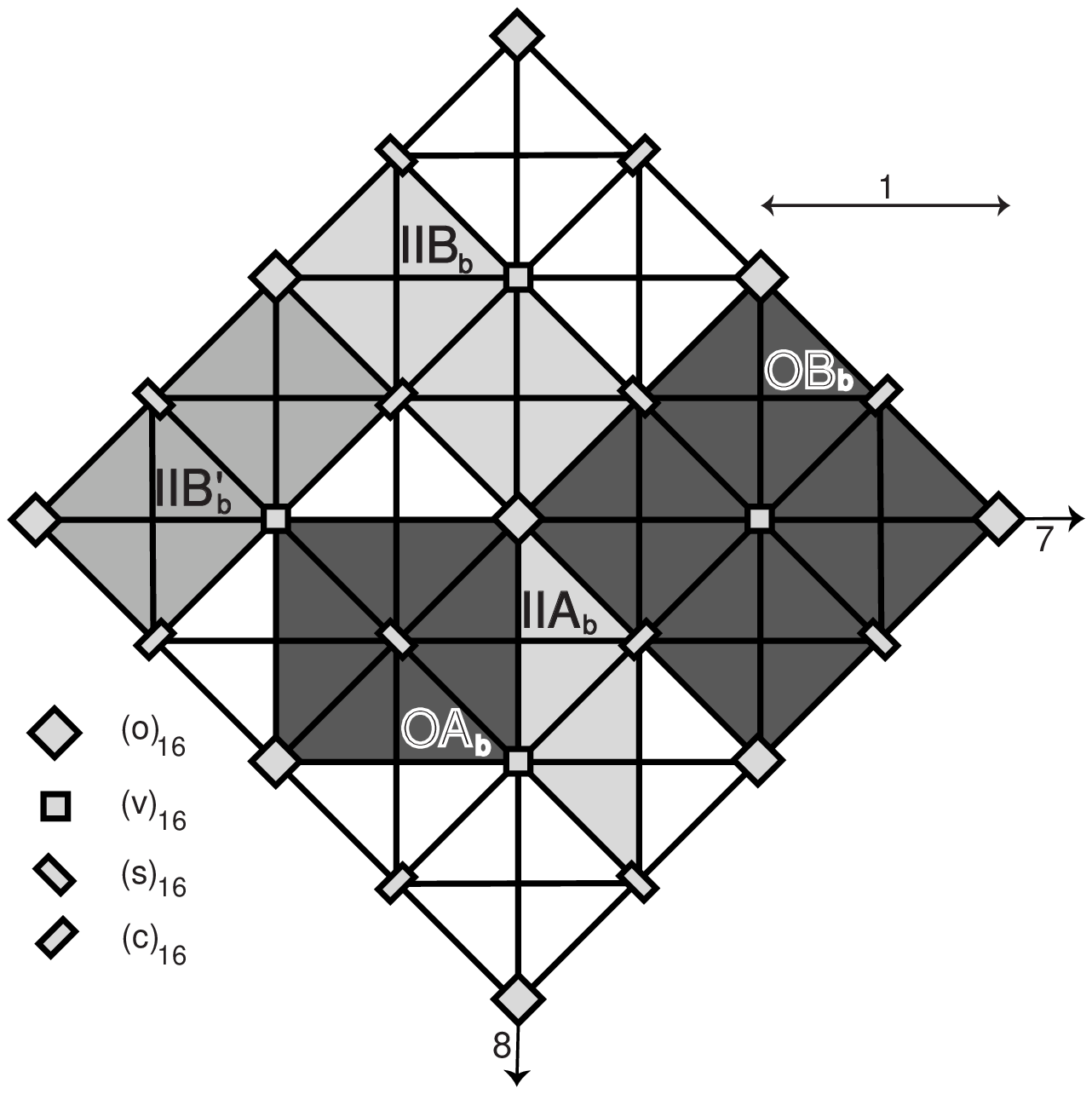} 
\begin{quote}
\baselineskip 14pt {\small Fig.2. Projected weight lattice of $SO(16)$ in the
$r_7 -r_8$ plane.  We see from Eqs.(\ref{ptorus}), (\ref{pbase}) that the
volumes
$\xi_i$ of the unit cells, exhibited in shaded areas, must be multiplied by
$(2\pi)^8. 2^{-8}$ to yield the 
$SO(16)$ compactification space torus volume of the  four bosonic theories  (in
units where
$\alpha^\prime =1/2$). The theories IIB and IIB' are isomorphic and differ by
the interchange of $s_{16}$ and $c_{16}$.  }\end{quote} 
The projection preserves the root length    and the
number of weight lattice points in each unit cell of the projected sublattices
$(r)_{16}$.  Hence it gives the correct  volumes of these unit
cells (although  the lengths of  spinor weights are not preserved). The
projected $SO(16)$ weight lattice is depicted in Fig.2.

The figure shows that the
unit $SO(16)$ cell volumes $\xi_i$ of the four bosonic theories are given by
\begin{equation}
\label{volume}
\xi_{OB_b}=2\ , \ \xi_{OA_b}=1\ , \ \xi_{IIB_b}=1\ ,\  \xi_{IIA_b}=1/2\, .
\end{equation} Restoring the $\alpha^\prime$ dependence and taking  into
account the contribution of the $E_8$ lattice,   one gets the 
  volumes $V_i$ of the configuration space tori   
\begin{equation}
\label{voltrue} V_i=(2\pi\alpha^{\prime\, 1/2})^{16}  2^{-8}\xi_i\, .
\end{equation}  These volumes  are thus related by 
\begin{equation}
\label{vall} V_{OB_b}= 2V_{OA_b}= 2 V_{IIB_b} = 4V_{IIA_b}.
\end{equation}

The   tori $\widetilde t$ of the four bosonic theories are, as group spaces, the  
maximal toroids $\widetilde{\cal T}/Z_c $ of the locally isomorphic groups 
$E_8\times \widetilde{SO}(16)/Z_c$ where $Z_c$ is a subgroup of the centre 
$Z_2\times Z_2$ of the
 universal covering group
$\widetilde{SO}(16)$.  We write
\begin{eqnarray}
\label{coset}\widetilde t\, ( OB_b)& = &\widetilde{\cal T}\, ,\nonumber\\
\widetilde t\, (OA_b)& =&
\widetilde{\cal T}/Z^d_2\, , \nonumber\\ \widetilde t\, (IIB_b)& =
&\widetilde{\cal T}/Z^+_2 ~~~{\rm or}~~~
\widetilde{\cal T}/Z^-_2\, , \nonumber\\ \widetilde t\, (IIA_b)& =&
\widetilde{\cal T}/(Z_2 \times Z_2) \, , \end{eqnarray} where
$ Z^d_2 = diag(Z_2 \times Z_2)$ and the superscripts $\pm$ label the two
isomorphic $IIB_b$ theories obtained by interchanging $(s)_{16}$ and
$(c)_{16}$. 

There is thus a unified picture for the four theories related to the global
properties of the $SO(16)$ group. The $OB_b$ theory built upon
$\widetilde{\cal T}$
 plays in some sense the r\^ole  of the `mother theory' of
the others.  One may view the different maximal toroids Eq.(\ref{coset}) as
resulting from the identification of  centre elements of 
$\widetilde{SO}(16)$, which are represented by   weight lattice points, with its
unit element. These identifications give rise  to the smaller shaded cells of
Fig.2.  In this way, the  unit cell of the
$IIB_b$ theory is obtained from the
$OB_b$ one by identifying the $(o)$ and $(s)$  lattice points\footnote{From
now on we drop the subscript 16 in the labeling of the lattice $(i)_{16}$.}  (or
alternatively the 
$(o)$ and the
$(c)$ lattice points) and therefore
 also the $(v)$ and $(c)$ lattice (or
 the $(v)$ and the $(s)$ lattice),  as seen in Fig.2. It is therefore equal
to the unit cell of the
$E_8$ lattice\footnote{The latter however does not contain  $(v)$ and $(c)$
lattice points.}.   The  unit cell of the
$OA_b$ theory is obtained by identification of $(o)$ and $(v)$, and  of
$(s)$ and $(c)$.

\vskip .5cm
\hskip .1cm\epsfbox{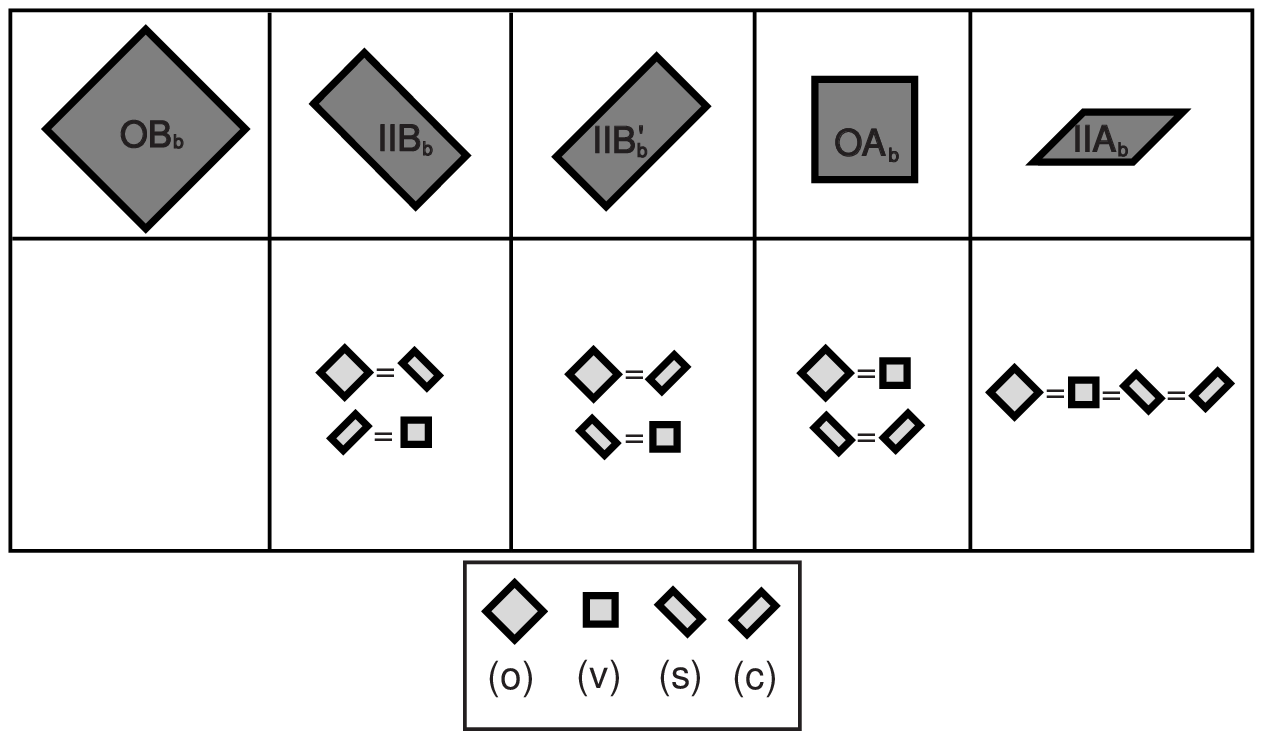}
\begin{quote}
\baselineskip 14pt {\small Fig.3.  Identifications of  centre elements of
$\widetilde{SO}(16)$ engendering the bosonic theories $IIB_b, \,
(IIB^\prime_b),\,  OA_b$ and $IIA_b $ from the mother  theory $OB_b$. The
dark shaded figures represent the tori of these bosonic theories in the
$SO(16)$ compact space. }
\end{quote} 
Finally,  the  unit cell of the
$IIA_b$ theory results from the identification of $(o), (v),(s)$ and $(c)$ and no
non-trivial centre element is left.  These results are summarized in Fig.3. 

We now evaluate the tension of the D9-branes. Tree amplitudes ${\cal
A}_{tree}$ are proportional to the square of the D-brane tension $T$. We can
 immediately deduce from Table I the following relations between the
tensions of the elementary  D9-branes of the different theories
\begin{equation}
\label{trel}  \sqrt{2}\, T_{OB_b}=T_{OA_b}=T_{IIB_b}=(1/\sqrt{2})\, T_{IIA_b}
\, .
\end{equation}
 To get their values, we recall that the tension 
$T^{bosonic}_{Dp}$ of a Dp-brane in the 26-dimensional uncompactified theory
is  \cite{pobb}
\begin{equation}
\label{dpb} T^{bosonic}_{Dp} = {\sqrt\pi\over2^4 
\kappa_{26}}(2\pi\alpha^\prime{}^{1/2})^{11-p}\, , \end{equation}
 where
$\kappa_{26}^2= 8\pi G_{26}$ and $G_{26}$ is the Newtonian constant in 26
dimensions. The tensions of the Dirichlet D9-branes of the four compactified
theories are obtained from Eq.(\ref{dpb}) by expressing
$\kappa_{26}$ in term of the 10-dimensional coupling constant
$\kappa_{10}$. Recalling that $\kappa_{26}=
\sqrt{V} \kappa_{10}$ where $V$ is the volume of the configuration space
torus, one finds,  using Eqs.(\ref{volume}) and (\ref{voltrue})
\begin{eqnarray}
\label{tob} T^{}_{OB_b} = {\sqrt\pi\over \sqrt{2}
\kappa_{10}}(2\pi\alpha^\prime{}^{1/2})^{-6}\, ,\\ \label{toa} T^{}_{OA_b} =
T^{}_{IIB_b} ={\sqrt\pi\over \
\kappa_{10}}(2\pi\alpha^\prime{}^{1/2})^{-6}\, ,\\ \label{t2a}
T^{}_{IIA_b}={\sqrt{2}
\sqrt\pi\over
\kappa_{10}}(2\pi\alpha^\prime{}^{1/2})^{-6}\, . \end{eqnarray} These are
consistent with  Eq.(\ref{trel}).

We now perform  the truncation Eq.(\ref{truncations}) on  the loop
amplitudes ${\cal A}$ listed in Table I. This yields the open string spectrum of
the fermionic D9-branes
\begin{eqnarray}
\label{FTOB} {\cal A}(OB_b)&\rightarrow& v_8 \equiv {\cal A}(OB)\, , \\
\label{FTOA} {\cal A}(OA_b)&\rightarrow& o_8+ v_8 \equiv {\cal A}(OA)\, ,\\
\label{FT2B} {\cal A}(IIB_b)&\rightarrow& v_8- s_8 \equiv {\cal A}(IIB)\, , \\
\label{FT2A} {\cal A}(IIA_b)&\rightarrow& o_8+ v_8-s_8-c_8\equiv {\cal
A}(IIA)\, .
\end{eqnarray}  These are indeed the amplitudes describing the fermionic
D9-branes of respectively $OB, OA, IIB$ and $IIA$.

 Furthermore tension is conserved in the truncation as proven in
reference~\cite{eht}. The tensions of the different bosonic D9-branes given in
Eqs.(\ref{tob})-(\ref{t2a}) are thus equal, when measured with the same
gravitational constant $\kappa_{10}$, to the  tensions of the corresponding
fermionic D9-branes \cite{pob,kets}. This is indeed a correct prediction. 

Up to now, we studied the properties of a single  `elementary'  D9-brane for 
the four $SO(16)$ bosonic strings and for the corresponding fermionic
theories. We now generalize the analysis to encompass several D-branes.
\vskip 1cm
\hskip 1cm \epsfbox{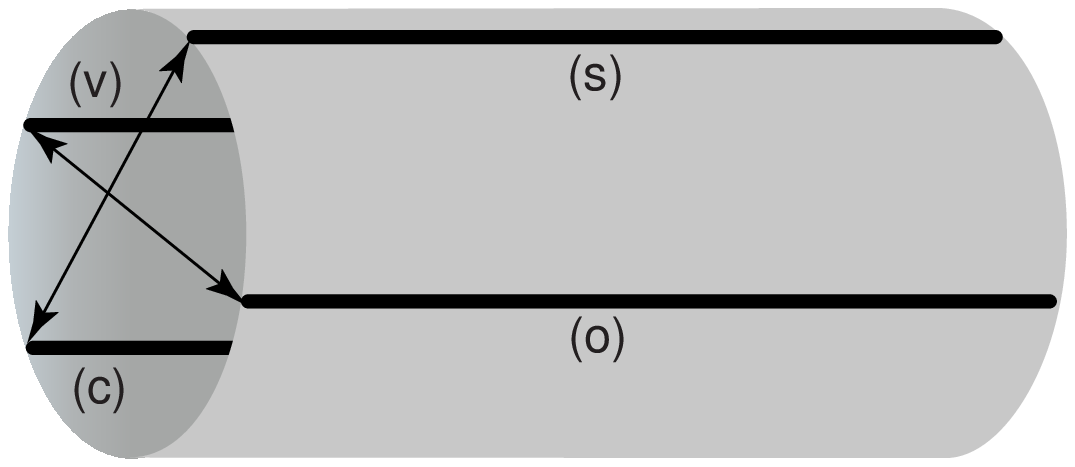}
\begin{quote}
\baselineskip 14pt {\small Fig.4.  The figure depicts the localization
of the $(o)\, $,  $(v)\, $, $(s)\, $, $(c)\,
$ D-branes  in the compact $SO(16)$ torus. The cylinder
represents the 9-dimensional  non-compact space and the circle the
8-dimensional compact $SO(16)$ torus. `Charge conjugate' branes are
connected by  arrowed lines.} 
\end{quote} 

\noindent
First, we remark that the relative position of the different
D9-branes in the eight compact dimensions of the $SO(16)$ torus, depicted in
Fig.4, is not arbitrary. Group symmetry requires that the partition function of
an open string with end points on D9-branes    be a linear combination of the
four $SO(16)$ characters. The vector {\bf d} separating the two points
where two distinct D9-branes meet the $SO(16)$ torus determines the
partition function of the  string starting at one point and ending at the other
after winding any number of times around the torus. The smallest eigenvalue
of the string Hamiltonian is $(1/2)\, {\bf d}.{\bf d}/\pi^2$. Therefore ${\bf d}/\pi$
must  be a weight and the D9-branes can only be separated in the compact
space (rescaled, as in Fig. 2, by a factor
$\pi^{-8}$) by a weight vector.  Consider for  instance two branes
 in the $OB_b$ theory, one located at $(o)$ and the other  located at $(v)$. The
partition function of a string beginning and ending on the same brane is
$o_{16}$, while  the partition function of an open string stretching between
them  is $v_{16}$.  For the other theories, the partition function of a string
beginning and ending on the same brane will then contain, in addition to
$o_{16}$, the characters corresponding to the strings stretched between
$(o)$ and all points identified with $(o)$.  This can be checked by comparing the
identifications indicated in Fig.3 with the partition functions $\cal A$ listed in
Table I. 

 If one chooses the location of  one elementary brane as the origin of the 
 weight lattice,  the other D9-branes can then only meet the $SO(16)$ torus
(rescaled by
$\pi^{-8}$) at a weight lattice point.  The number of distinct elementary branes
is, for each of the four bosonic theories, equal to the number of distinct
weight lattice points in the unit cell. For the mother theory
$OB_b$  there are four possible  elementary D9-branes. We label them by their
positions in the unit cell, namely by
$ (o), (v), (s)$  and $(c)$. Note that these weight lattice points  represent the
centre elements of the $\widetilde{SO}(16)$. 
 For the other theories the unit cells are smaller and there are fewer
possibilities. The unit cell of the
$IIB_b$ theory  allows  only  for  two distinct branes  $ (o)= (s)$,  $ (c)= (v)$, as
seen from Fig.3.  For $IIB_b^\prime$, one simply interchanges $(s)$ and $(c)$. 
Similarly for the $OA_b$ theory, we have the two branes
$(o)=(v)$ and
$(s)=(c)$, and finally for the `smallest' theory $IIA_b$, we have only one
elementary brane $(o)=(v)=(s)=(c)$.

It is interesting to note that the charge conjugation of the truncated fermionic
strings is encoded in their bosonic parents. A brane sitting at
$(v)$ can always be joined by an open string to a brane sitting at $(o)$. The
partition function of such a string is given by the character $v_{16}$ and
therefore the two branes can exchange closed strings with tree amplitude 
${\cal A}_{tree} = o_{16}+ v_{16}-s_{16}-c_{16} $ as seen from the
S-transformation Eq.(\ref{sixteen}).  Namely the closed string exchange  
describing the interaction between these two
 branes has opposite sign for the $(s)$ and $(c)$ contribution  as compared to
 the closed string exchange between D9-branes located at the same point.
Using Eq.(\ref{truncations}), we see that this shift of sign persists  in the
fermionic theories where the  above tree amplitude becomes
$o_8+ v_8+s_8+c_8 $. This shift of sign  is the  RR-charge conjugation between
fermionic D9-branes.  It is encoded in the bosonic string as a  shift by the
lattice vector $(v)$  (see Fig.4). In particular, when $(o)$ and $(v)$ are
identified, all branes of the fermionic  offsprings are neutral. These are always
unstable branes, as the truncation of $v_{16}$ is
$o_8$ and contains a tachyon. Charged branes are always stable.

The distinct fermionic D9-branes and their charge conjugation properties can
thus be directly read off from  Fig.2. They are summarized in Table II,
where the charge is indicated by a superscript $+,-$ or $0$, and additional
quantum numbers by a subscript.

\vskip .8cm
\begin{center}
\begin{tabular}{||c||c|c|c||}
\hline &${\cal A}^{trunc}$&Fermionic D9-branes&stability\\
\hline\hline
$OB_b\to OB$&$v_8$&$D^+_1+D^+_2+D^-_1+D^-_1$&stable\\
\hline
$OA_b\to OA$&$v_8+o_8$& $D^0_1+D^0_2$&unstable\\
\hline
$IIB_b\to IIB$&$v_8-s_8$& $D^+ +D^-$&stable\\
\hline
$IIA_b\to IIA$&$v_8+o_8-s_8-c_8$& $D^0$&unstable\\
\hline
\end{tabular}
\vskip .5cm
 Table II
\end{center}
\vskip .1cm

Having discussed the elementary space-filling D9-branes of the different
bosonic theories and their fermionic offsprings we now turn to the discussion
of Chan-Paton multiplicities and the general annulus amplitudes of each
bosonic theory.

The general direct annulus amplitude of the mother $OB_b$ theory can be
written as
\begin{equation}
\label{genob} {\cal A} = \sum_{l,m,i} {\cal A}^i_{\, \, lm}\, n^l {\bar n}^m\,
i_{16}\, ,
\end{equation}
 where the $n^l,\bar n^m$ are the Chan-Paton multiplicities (one for each end
of the open strings) and the ${\cal A}^i_{\, \, lm}$ are coefficients. The sum is
over the lattice points  $(l), (m)$  in the unit cell  and over the characters
$i_{16}$; $l, m,i = o,v,s,c$. As seen from Fig.2,  a string stretched between
branes located at  lattice points 
$(l), (m)$ has end points separated by a weight vector equal to a vector
joining  $ (o)$ to some lattice point $ (i)$. Its partition function is
$i_{16}$. Hence,  the coefficients ${\cal A}^i_{\, \, lm}$ are given by the
 multiplication table of the conjugacy classes or, equivalently,  by the fusion
rules Eq.(\ref{bfusion}). Writing these fusion rules as
\begin{equation}
\label{gfusion}
\, [i] [j] = \sum_{k} {\cal N}_{ij}^{\, \, k} \, [k]\, , 
\end{equation}  where the ${\cal N}^{ij}_{\, \, k}$ are the integer fusion-rule
coefficients we have
\begin{equation}
\label{cardy} {\cal A}^k_{\, \, ij}= {\cal N}_{ij}^{\, \, k}\, . 
\end{equation} We recover, in this particular case, the  Cardy
solution for the annulus for diagonal conformal field theories \cite{cardy,
sagn1}.

 Explicitly, the general $OB_b$ amplitude Eq.(\ref{genob}) takes the form
\begin{eqnarray}
\label{cpob} {\cal A}(OB_b)&=& (n_o{\bar n}_o +n_v{\bar n}_v+ n_s{\bar n}_s
+n_c{\bar n}_c) \, o_{16}\nonumber\\ &+& (n_o{\bar n}_v+ n_v{\bar n}_o+
n_s{\bar n}_c+ n_c{\bar n}_s) \, v_{16} \nonumber\\ & +& ( n_o{\bar
n}_s+n_s{\bar n}_o+ n_v{\bar n}_c+n_c{\bar n}_v) \, s_{16}\nonumber \\ &+&
(n_o{\bar n}_c+n_c{\bar n}_o+n_v{\bar n}_s+n_s{\bar n}_v)\,  c_{16}\, .
\end{eqnarray} Performing the identifications depicted in Fig.3, we
immediately get from Eq.(\ref{cpob})  the amplitudes for the other theories
\begin{eqnarray}
\label{cpoa} {\cal A}(OA_b)&=& (n_o{\bar n}_o + n_s{\bar n}_s) \, (o_{16} +
v_{16})\nonumber\\ & +& ( n_o{\bar n}_s+n_s{\bar n}_o) \, (s_{16}+ c_{16})\, ,
\\
\nonumber\\
\label{cp2b} {\cal A}(IIB_b)&=& (n_o{\bar n}_o + n_v{\bar n}_v) \, (o_{16} +
s_{16})
\nonumber\\ & +& ( n_o{\bar n}_v+n_v{\bar n}_o) \, (v_{16}+ c_{16})\, ,\\
\nonumber\\
\label{cp2a} {\cal A}(IIA_b)&=& n_o{\bar n}_o \, (o_{16} +v_{16}+s_{16}+
c_{16})\, .
\end{eqnarray}  Note that in the identification, care has been exercised not to
overcount the branes. Namely in going from Eq.(\ref{cpob}) to Eq.(\ref{cpoa})
by identifying $(v),(c)$ to $(o), (s)$, one must put $n_v$ and
$n_c$ equal to zero as these branes have already been counted in $n_o$ and
$n_s$. Similarly, one  put $n_s$ and $n_c$   to zero in Eq.(\ref{cp2b}) and 
$n_v,n_s$ and $n_c$ to zero in Eq.(\ref{cp2a}).

These results may be viewed as particular realizations of boundary conformal
field theory. Eqs.(\ref{cpob}) - (\ref{cp2a}) have the  structure generalizing
Eq.(\ref{genob})

\begin{equation}
\label{geneam} {\cal A} = 
\sum_{a,b,i} {\cal A}^i_{\, \, ab} n^a {\bar n}^b\, i_{16}\, .
\end{equation}  The indices $a$ and $b$ are `charge' indices  representing the
branes where the open string may end, $i$ labels  the  characters and
the
${\cal A}^i_{\,
\, ab}$ are integer-valued coefficients  \cite{sagr}.  In the $OB_b$ case, as
discussed above, (see also Fig.2) there are four possible locations for the
open string end points and the charge indices
$a$ and $b$ take the same four possible values $o,v,s$ and $c$ as  
$i$.  For the
three other models we have less charge indices than character labels.  
Following the discussion on the possible location of the D9-branes above, the
$OA_b$ theory and the
$IIB_b$  have  both  two charge indices, respectively $a=o,s$ and
$a=o,v$. The
$IIA_b$ theory has only one possible charge label $a=o$. 

Finally, the general amplitudes of the fermionic offsprings are obtained from
Eqs.(\ref{cpob}) - (\ref{cp2a}) by the  truncation Eq.(\ref{truncations}).

\setcounter{equation}{0}
\section{The  even D-brane paradox}

In this section we discuss the truncation of bosonic D-branes to lower 
dimensional fermionic Dp-branes (p$<$9). This is a non-trivial problem for  the
following reason. In fermionic string theories, a T-duality  interchanges type
$IIA$ with $IIB$, and type $OA$ with $OB$ while transmuting  D9-branes to 
D8-branes without changing their corresponding ${\cal A}^{trunc}$
amplitudes\footnote{The only change in the integrand of the amplitudes is the
power law of the moduli which, along with the Dedekind functions, are omitted
throughout the paper.} given in Table~II, although the latter may be
conveniently rewritten in terms of representations of lower dimensional
orthogonal groups
\cite{sagd}. This interchange is generally viewed as the compatibility of
T-duality with world-sheet  supersymmetry. It can also be understood  in the
light-cone gauge, or equivalently in the truncated theory, as the consistency
of T-duality with the closure of the  algebra
$so_{diag}(8) ={\rm diag} [so_{trans}(8) \times so_{int}(8)]$ extended to the
Lorentz algebra $so\, (9,1)$ (see Appendix A). 

Therefore, {\em in any given
fermionic  theory}, the amplitudes of Dp-branes {\em whose $p$  have a
definite parity}  are essentially the same (see footnote 8), but
differ from the amplitudes of Dp-branes of the other parity.

This is in contrast with parent bosonic theories, where in a given theory
${\cal A}_p={\cal A}_{p-1}$ for $p \le 9$. Bosonic D8-branes are thus, as the D9
branes,  uniquely determined by imposing Dirichlet boundary 
conditions in the 16-dimensional compact space. So, if we were to obtain
a  fermionic D8-brane\footnote{The argument is easily generalized to any 
even-dimensional brane.} from the truncation ${\cal A}^{trunc}_8$ of a D8-brane
in the bosonic  parent
theory, we would get a wrong result. The 
loop amplitude of a fermionic D8-brane, ${\cal A}^{\prime\, trunc}_8$ is
different from  ${\cal A}^{trunc}_8$, as shown in Table III.
\vskip .5 cm
\begin{center}
\begin{tabular}{||c||c|c||c|c||}
\hline 
&
${\cal A}_9={\cal A}_8$&
${\cal A}^{trunc}_8$&
& 
${\cal A}^{\prime\, trunc}_8$\\
\hline\hline
$OB_b$&$o_{16}$&$v_{8}$&$OB$&$o_{8}+v_{8}$\\
\hline
$OA_b$& $o_{16}+ v_{16}$&$o_{8}+v_{8}$&$OA$&$v_{8}$\\
\hline
$IIB_b$& $o_{16}+ s_{16}$&$v_{8}-s_{8}$&$IIB$&$o_{8}+ v_{8}- s_{8}-c_{8}$\\
\hline
$IIA_b$&$o_{16}+ v_{16}+ s_{16}+ c_{16}$&$o_{8}+v_{8}- s_{8}-c_{8}$&$IIA$&
$v_{8}-s_{8}$\\
\hline
\end{tabular}
\vskip .5cm
 Table III
\end{center}

The puzzle is twofold. On the one hand, we appear to lack, in  a {\em given
bosonic parent theory}, a whole family of branes with $p+1$ Neumann boundary
conditions in the ten non-compact space-time dimensions  whose lattice
partition function
 coincides with the D9 lattice partition function of a  {\em
different} parent bosonic theory, namely the one obtained by
interchanging
$OB_b
\Leftrightarrow OA_b$ and $IIB_b \Leftrightarrow IIA_b\,$!  On the other hand,
we get irrelevant D-branes for $p$ even.

The fate of the irrelevant branes  is easily understood  from  the fact
that T-duality in the bosonic string does not change the bosonic parent
theory. As
${\cal A}_8$ can be obtained from
${\cal A}_9$  from such T-duality, it is clear   that a truncation leading
to
${\cal A}^{trunc}_8$ in Table III would be inconsistent with the Lorentz
invariance of the truncated theory. Hence D8-branes, and more generally
Dp-branes with $p$ even disappear from the fermionic subspace.
\vskip .5  cm
\begin{center}
\begin{tabular}{||c||c|c||}
\hline &${\cal A}_p\, ,~~p$  odd&${\cal A}^\prime_p\, , ~~p$  even\\
\hline\hline
$OB_b$&$o_{16}$&$o_{16}+ v_{16}$\\
\hline
$OA_b$& $o_{16}+ v_{16}$&$o_{16}$\\
\hline
$IIB_b$& $o_{16}+ s_{16}$&$o_{16}+ v_{16}+ s_{16}+ c_{16}$\\
\hline
$IIA_b$&$o_{16}+ v_{16}+ s_{16}+ c_{16}$&$o_{16}+ s_{16}$\\
\hline
\end{tabular}
\vskip .5cm
 Table IV 
\end{center}
The `missing' loop amplitudes ${\cal A}^\prime_p$ are shown in Table IV.  
We shall show in the next section that the
$p$-even amplitudes ${\cal A}^\prime_p$ are obtained from bosonic
D($p$+8)-branes, namely from branes which wrap once around the $SO(16)$
compact dimensions.  
On the other  hand,
the bosonic origin of the $p$-odd amplitudes is clear.  They arise from the
truncation of bosonic Dp-branes with the same lattice partition function as the
bosonic  D9-branes. It will be
seen below that the correspondence between the $p$-odd and the $p $-even
amplitudes, exhibited in Table IV,  is rooted in the existence of `odd'
E-dualities  which interchange simultaneously the  $OB_b$ and $OA_b$ (or
$IIB_b$ and $IIA_b\,$) theories, and the Dirichlet branes with the wrapped ones
in the $SO(16)$ compact dimensions.

This solves   the above puzzle.  Completion of
the T-duality in  ten dimensions by an  `odd' E-duality in the eight
dimensions compactified on an
$SO(16)$ torus  yields  all the bosonic branes listed in Table IV
from the D9-branes of Table I  and ensures  the Lorentz invariance of
the fermionic subspaces.  This will be made explicit in the next section.
\setcounter{equation}{0}
\section{Even and odd  E-dualities} 

In the action formalism, toroidal compactifications are described by the
action Eq.(\ref{action}) in the background  metric $g_{ab}$ and antisymmetric
tensor
$b_{ab}$. The left and right momenta of closed strings are given by
Eq.(\ref{close}) which can be rewritten as
\begin{eqnarray} {\bf p_R} &=& {1\over2} m_a {\bf e^a} + n^a e_{ab}{\bf e^b}\,
,   \nonumber \\ {\bf p_L} &=& {\bf p_R} - 2n^a {\bf e_a}\ , \label{eclose}
\end{eqnarray} where $\{{\bf e_a}\}$ defines the configuration space torus
(see Eq.(\ref{ptorus})) and $\{{\bf e^a}\}$ is the lattice-dual basis.  The metric
$g_{ab}$ is given by Eq.(\ref{pmetric}) and we have defined
\begin{equation}
\label{edef} e_{ab}=g_{ab}+b_{ab}\, .
\end{equation}  Here again, we  concentrate on the $SO(16)$ contribution of 
$E_8 \times SO(16)$.  Therefore, from now on, the indices $a,b$ run from 1 to
8.   Indeed, the $E_8$ lattice partition function will not  relevant,  except for a
subtlety in the open sector to be discussed at the end of this section.

In the case of EN lattices and in particular for the
$OB_b$ and
$IIB_b$ theories, there is a well-known Lagrangian realization \cite{egrs} 
which we shall refer to as   the `standard' realization. It is achieved by taking, 
in accordance with Eq.(\ref{pbase}), 
${\bf w_a}= 2 {\bf e_a}= {\bf r_a}$ with ${\bf r_a}$  the simple roots, and by
choosing the following `canonical' constant antisymmetric background field
\begin{equation} b_{ab} = +{\bf e_a.e_b}\ \hbox{for}\ a> b\ , \ = -{\bf e_a.e_b}\
\hbox{for}\ a< b\ ;
\quad b_{ab} =0\ \hbox{for}\ a=b\, .
\label{oldb}
\end{equation} This is not the only choice.  Vectors ${\bf w_a}$ spanning
different unit cells of the lattice along with canonical or non-canonical
$b_{ab}$ are also possible. Here we shall restrict, for the four bosonic
theories,  the Lagrangian description to canonical $b_{ab}$ tensors  given by
Eqs.(\ref{oldb}), whether or not ${2\bf e_a}$ is a root.  In Appendix~B, we
present explicitly such  canonical realizations.
The corresponding sets of basis vectors
$\{{\bf w_a}\}_r \,
\, r=1,..,4$ given there span the four sublattices $(r)$ of Eqs.(\ref{sublattice}).
Note that for the two EN cases 
$OB_b$ and $IIB_b$, these realizations are not standard.
 Indeed the unit cell chosen is not spanned by simple roots and furthermore
one of the vector is not a root and  has
${\bf w^2}=4$ (instead of ${\bf w^2}=2$). This fact will be an important 
ingredient when solving the dilemma of the previous section.

Consider E-duality transformations \cite{gpr}. On $e_{ab}$, it is
defined as follows
\begin{equation}
\label{edual} e \rightarrow {e^{-1} \over 4}\, ,
\end{equation} with
\begin{equation}
\label{gbdual} ({e^{-1} \over 4})^{ab}=G^{ab}+B^{ab}.
\end{equation}
When the transformation Eqs.(\ref{edual}),(\ref{gbdual}) is performed, the
Hamiltonian of  the closed string sector is invariant under the combined
exchange of the background fields $g_{ab}$ and $b_{ab} $ with the E-dual metric
$G^{ab}$ and antisymmetric tensor
$B^{ab}$, and of winding modes with momenta,
$m_a \leftrightarrow n^a$. 

In the open string sectors  E-duality interchanges \cite{ers} Dirichlet
boundary conditions given by Eq.(\ref{diro})  (or expressed in the closed
string tree channel by Eq.(\ref{dirc})~) with generalized Neumann boundary
conditions defined  by
\begin{equation}
\label{neumann} [g_{ab}\partial_\sigma X^b - b_{ab}
\partial_\tau X^b]
\vert^{\sigma=\pi}_{\sigma=0} =0 \, .
\end{equation}  In the tree channel this condition becomes 
\begin{eqnarray}
\label{neumannt} g_{ab}\partial_\tau X^b - b_{ab}
\partial_\sigma X^b &=&0\, ,\\
\hbox{or} \qquad\qquad\qquad m_a &=&0\, .
\end{eqnarray}  Thus E-duality maps  a D-brane localized on the torus onto a
D-brane completely wrapped on  it or vice-versa. 

We now analyse  the effect of the E-duality on the left and right momenta
Eq.(\ref{eclose}). Let  $\{{\bf E^a}\}$ be a basis of the E-dual lattice. We have
$G^{ab}={\bf E^a.E^b}$. We  denote by $\{{\bf E_a}\}$ its lattice-dual basis 
defined by ${\bf E_a.E^b}=\delta_a^b$. E-duality transforms ${\bf p_R}$ and
${\bf p_L}$ according to
\begin{eqnarray}
\label{edualbis} {\bf e_a}& \rightarrow & {\bf E^a}\, , \nonumber\\ {\bf e^a}& 
\rightarrow & {\bf E_a}\, ,\nonumber\\ m_a &\leftrightarrow &n^a\, . 
\end{eqnarray} We may choose the basis ${\bf E_a}$ in such a way that the
E-dual right   momenta ${\bf p^D_R}$ are equal to ${\bf p_R}$ given by
Eq.(\ref{eclose}). We get,
 \begin{equation}
\label{dualbasis} {\bf E_a}=2 e_{ab} {\bf e^b}.
\end{equation}   Using Eqs.(\ref{edualbis}) and (\ref{dualbasis}) we have for 
the E-dual right and left momenta ${\bf p^D_R}$ and ${\bf p^D_L}={\bf
(p^D_L)}_a {\bf e^a}$ 
\begin{eqnarray}
\label{dmomenta} {\bf p^D_R} &=& {\bf p_R}\, , \nonumber \\ {\bf (p^D_L)}_a
&=& -(e^{t} . e^{-1})_a^{\, \, b} {\bf (p_L)}_b\, , \end{eqnarray} where the
transpose matrix
$e^{t}_{ab}=g_{ab}-b_{ab}$. The left momenta are rotated under E-duality
\cite{ers,hkms}.  Indeed, the transformation Eq.(\ref{dmomenta}) is
represented in the Cartesian basis by a rotation matrix $ {\cal R}_i^{\, \, j}$:
\begin{equation}
\label{rotation} {\bf (p^D_L)}_i = {\cal R}_i^{\, \, j} {\bf (p_L)}_j \, \,
\rm{with}
\, \,  {\cal R}_i^{\, \, j}= - {\bf e^a}_i(e^{t}.e^{-1})_a^{\, \, b} {\bf e_b}^j\, .
\end{equation} 

The effect of this rotation is interesting. One might have thought that, at an
enhanced symmetry point, E-duality always maps the  theory onto itself.
However this need not be the case if the closed string spectrum can be
mapped onto a spectrum of a different theory with the same degeneracy. This
is the case for the $OA_b$ and $OB_b$ spectra or for the
$IIA_b$ and $IIB_b$ spectra. In fact, starting from any  Lagrangian realization
of $OA_b$ and $IIA_b$,  E-duality {\em always} maps  $OA_b$ onto
$OB_b$ and $IIA_b$ onto $IIB_b$. The demonstration of this theorem is
presented in Appendix C. We shall characterize such an E-duality as  `odd' to
distinguish it from  the `even' E-dualities which map a theory onto itself.    
E-duality applied to
$OA_b$ and $IIA_b$ theories is thus always odd.

On the other hand starting with $OB_b$ or $IIB_b$ one may have both even and
odd E-dualities. For instance, $OB_b$ and $IIB_b$ in their standard  Lagrangian
realization are mapped onto themselves by even E-duality, but the inverse of
an E-duality mapping $OA_b$ (resp. $IIA_b$) onto $OB_b$  (resp. $IIB_b$)  is of
course odd. Realizations connected by odd  (resp. even) E-dualities are called
odd (resp. even). Examples of odd realizations are given in Appendix B.

One can understand the origin of odd E-dualities in the following way. It is
always possible to construct a unit cell of the $OB_b$ or 
$IIB_b$ lattice, which contains a single vector ${\bf w_a}$ of squared length 4
 and to deduce from it a canonical Lagrangian realization (see Appendix~B).
Taking the determinant of Eq.(\ref{dualbasis}), we get the
 unit cell volume $V_{dual}$ of the E-dual theory in terms of the unit cell 
volume $V$  of the original theory
\begin{equation}
\label{relvol} V_{dual}^{-1}= 2^8 ({\rm det}\, e) \, \, V^{-1}\, .
\end{equation} Since we are dealing with canonical realizations,
$e_{ab}$ is a triangular matrix. Henceforth, \begin{equation}
\label{vcondbis} {\rm det}\, e = \prod_{a=1}^8 {\bf e_a. e_a} =2^{-16}
\prod_{a=1}^8 {\bf w_a. w_a},
\end{equation} where ${\bf w_a}$ define a unit cell of
$OB_b$ or
$IIB_b$ (modulo the `dummy' $E_8$). If all the ${\bf w_a}$ are roots of  $SO(16)$
or
$E_8$, we get
$V_{dual}=V$.  On the other hand, if one takes a unit cell of the root lattice
containing one vector of squared length four, we get
$V_{dual}=V/2$. This agrees with an E-duality transform of the volume of a
unit cell from
$OB_b$ to that of $OA_b$, and similarly, from $IIB_b$ to
$IIA_b$. Such a construction of odd E-dualities is exhibited in Appendix B.

We are now in a position to discuss the existence of the different bosonic
Dp-branes given in Table IV. 

We take as input the bosonic D9-branes of Table I which are uniquely defined
independently of the Lagrangian realization. The Dp-brane spectra are then
fully determined by considering  {\it odd} realizations of the four theories. 
The lower dimensional ones are 
 univocally obtained by performing together  odd E-dualities and
T-dualities, as required by the Lorentz invariance of the truncated theories.
Consider indeed  the
$OA_b$ and $IIA_b$ theories, which only admit odd realizations.  Their 
amplitudes ${\cal A}_{p}$ for odd $p$ are obtained from  Dirichlet boundary
conditions in the compact space while the  amplitudes  ${\cal A^\prime}_{p}$
for even
$p$ describe D(p+8)-branes wrapped in the
$SO(16)$  compact space. The latter are obtained from an odd realization of
$OB_b$ and
$IIB_b$ by performing an E-duality on the Dp-branes of these two
theories.  Similarly, even $p$ amplitudes of the $OB_b$ and
$IIB_b$ theories are given by wrapped branes which are the E-dual of Dirichlet
branes of $OA_b$ and $IIA_b$.

The $OB_b$ and $IIB_b$ also have  even realizations in which  the
Neumann boundary conditions in compact space simply duplicate the spectra
of the Dirichlet conditions. Hence they would only  generate the odd
amplitudes ${\cal A}_{p}$ while the odd realizations generate all of them. The
richer open spectra of the odd realizations  point towards their 
fundamental character.

Up to now we have considered Dirichlet boundary conditions on the
 $E_8$ torus. One may also consider Neumann boundary
conditions, that is D-branes wrapped on this 
torus.  The corresponding tensors $g_{ab}$ and $b_{ab}$ are equivalent to
those defined on the  $SO(16)$ torus for  the $IIB$ theory and thus 
compactifications on the $E_8$ torus admit even and odd realizations. Both 
have 
${\cal G} =E_8\times E_8$ symmetry in the closed string sectors but this
symmetry is broken in odd realizations for  open strings  with Neumann
boundary conditions. In fact the truncation of the  amplitudes in this case is
inconsistent, and for Neumann boundary conditions one must take an even
realization. To understand this point, we first notice that  the truncation of
$E_8$ is of a different nature than the truncation of $SO^{\, \prime\,}(8)$. In the
former case, one indeed drops, in addition to the oscillator states, {\em all}
momentum states. Thus the only remnant of $E_8$ in the truncated theory is
the coefficient multiplying the $o_{16}$ character in the $E_8$ lattice
contribution to the annulus amplitude. The latter appears, for Dirichlet
boundary conditions, in the third row of Table I, both in the tree and in the loop
expression for this amplitude. For even realizations of $E_8$, these correct
coefficients are not altered when considering Neumann branes, as their lattice
partition functions are the same as the Dirichlet ones. For odd realizations, the
Neumann amplitude (or equivalently the Dirichlet amplitude of the E-dual
theory) appears in the fourth row of the same table. We see immediately that,
while the coefficient of the truncated theory in the loop expression of the
annulus amplitude is correct, the coefficient is off by a factor of 2 when
expressed in the tree channel. This translates the fact that for odd
realizations, truncation of the
$E_8$ contribution does not commute with the S-modular transformation in
the open string sectors.

We  see that, if we choose for all four theories the D9-branes to be those
uniquely determined by the Dirichlet conditions in compact space, consistency
of the truncation with Lorentz and modular invariance fully determines, for all
lower dimensional fermionic branes, the amplitudes of the bosonic parents to
be those given in Table IV. The crucial element which led to this astonishing
result, is the existence  of odd E-dualities. These reflect  a hidden link
between the fermionic theories, expressed in the bosonic string through the
global group properties of 
$\widetilde{SO}(16)$. 

\setcounter{equation}{0}
\section{Tadpole-free descendants and truncation} 

In this section we describe the open descendants of the closed bosonic
theories. These are determined by imposing the {\it tadpole condition}
 \cite{sagr} on the bosonic string, namely by
imposing that divergences due to massless tadpoles cancel in the 
vacuum amplitudes. We will show that the bosonic $OB_b$,
$IIB_b$ and
$OA_b$ theories admit tadpole-free open bosonic descendants and that those
descendants give after truncation the three open fermionic string
theories which are anomaly or tadpole-free. In our previous paper
\cite{eht}, we already derived the open descendants of $OB_b$ and $IIB_b$
which, after truncation, give respectively the tadpole-free
$[SO(32-n)
\times SO(n)]^2$ theory and the anomaly free
$SO(32)$ type I theory. These results were derived for $OB_b$ and
$IIB_b$ from the annulus  ${\cal A}$ and Klein bottle ${\cal K}$ amplitudes
describing wrapped D25-branes and orientifolds. We shall keep this approach
here for the orientifolds, but in accordance with the method used in the
previous sections, we consider Dirichlet D9-branes. This approach  is
equivalent to the one of reference \cite{eht} for the $OB_b$ and
$IIB_b$ theories\footnote{The amplitudes
${\cal A}$ used in reference \cite{eht} for $OB_b$ and $IIB_b$ describing
D25-branes have the same  lattice contribution as the D9-brane amplitudes
used here and  they can be interpreted as the E-dual of the latter in an even
realization of the $OB_b$ and
$IIB_b$ theories.}, but will allow the construction of the open descendant of
$OA_b$ which was not obtained there.

A first step in obtaining the open descendant corresponding to the four
bosonic string theories characterized by the tori amplitudes ${\cal T}$
Eqs.(\ref{BOB})-(\ref{B2A}) is the construction of the  Klein bottle amplitudes
${\cal K}$. These are obtained from the amplitudes
${\cal T}/2+ {\cal K}$, which are the torus closed string partition functions
${\cal T}$ with the projection operator
$(1+\Omega)/2$ inserted, where $\Omega$ interchanges the left and right
sectors:
$\Omega |L,R> =|R,L>$.  This can be done for $OB_b$, $IIB_b$ and $OA_b$ but not
for
$IIA_b$, because $\Omega$ in that case is not a symmetry of the theory. The
$IIA_b$ theory does not admit any open descendant. As a consistency check we
 note that the realization of $OA_b$ is the only one  presented in
Appendix B which does not fulfill the condition $4b_{ab} \in {\cal Z}$
ensuring that
$\Omega$ is a symmetry of a toroidally compactified theory~\cite{sagn2}. The
projection on  $\Omega$ eigenstates amounts  to impose the
 condition 
\begin{equation}
\label{kcon1} {\bf p_R = p_L}\, ,
\end{equation}   on the closed string momenta Eqs.(\ref{close}). Acting with
$\Omega /2$ on the three different tori Eqs.(\ref{BOB})-Eqs.(\ref{B2B}), one
finds the three Klein bottle amplitudes\footnote{Recall that we display only 
the
$SO(16)$ contribution  of the amplitudes.}
\begin{eqnarray}
\label{KOB} {\cal K}(OB_b) & = & {1 \over 2} (o_{16}+v_{16}+s_{16}+c_{16})\, ,\\
\label{K2B} {\cal K}(IIB_b) & = & {1 \over 2} (o_{16}+s_{16})\, ,\\ \label{KOA}
{\cal K}(OA_b) & = & {1 \over 2} (o_{16}+v_{16})\, .
\end{eqnarray}

The two remaining amplitudes with vanishing Euler characteristic,   the annulus
${\cal A}$ and the M\"obius strip
${\cal M}$, determine the open string partition function.  The
 annulus amplitudes of  D9-branes with generic Chan-Paton
multiplicities are given in Eqs.(\ref{cpob}), (\ref{cp2b}) and (\ref{cpoa}). For
the unoriented string considered here, $n=\bar n$ and the  amplitudes 
 must be divided by two. Denoting these   amplitudes by ${\cal A}^{un}$, we have
\begin{eqnarray}
\label{cuob} {\cal A}^{un}(OB_b)&=& {1\over 2} (n_o^2 +n_v^2 + n_s^2 +n_c^2)
\, o_{16} + ( n_o n_v + n_s n_c) \, v_{16} \nonumber\\ & + & ( n_o n_s+ n_v n_c)
\, s_{16} +(n_o n_c + n_v n_s) c_{16}\, , \\
\label{cuoa} {\cal A}^{un}(OA_b)&=& {1\over 2} (n_o^2 + n_s^2) \, (o_{16} +
v_{16})  + n_o n_s \, (s_{16}+ c_{16})\, , \\
\label{cu2b} {\cal A}^{un} (IIB_b)&=& {1\over 2} (n_o^2 + n_v^2) \, (o_{16} +
s_{16}) + n_o n_v \, (v_{16}+ c_{16})\, .
\end{eqnarray}

To get the M\"obius amplitudes  ${\cal M}$ and to implement the tadpole
condition we express the Klein bottle  and annulus amplitudes
Eqs.(\ref{KOB})-(\ref{KOA}) and  Eqs.(\ref{cuob})-(\ref{cu2b}) as
 closed string tree  channel amplitudes. Using the S-transformation of the
characters Eq.(\ref{sixteen}) and the scaling of the moduli in the integration
variables one finds
\begin{eqnarray}
\label{TKOB} {\cal K}_{tree}(OB_b) & = & 2^5 \, o_{16}\, ,\\ \label{TUOB} {\cal
A}^{un} _{tree}(OB_b)& =& 2^{-7} \: [ (n_o +n_v + n_s +n_c)^2 \, o_{16} 
\nonumber \\ & + & ( n_o+ n_v - n_s- n_c)^2 \, v_{16}\ \nonumber\\ & + & (
n_o-n_v + n_s- n_c)^2 \, s_{16}\nonumber\\ & + & (n_o- n_v - n_s+ n_c)
c_{16} ] \, ,\\ \label{TK2B} {\cal K}_{tree}(IIB_b) & = & 2^4 \, (o_{16}+s_{16})\,
,\\
\label{TU2B} {\cal A}^{un} _{tree}(IIB_b)&=& 2^{-6} [ (n_o + n_v)^2 \, o_{16} +
(n_o -n_v)^2 \, s_{16} ]\, ,\\
\label{TKOA} {\cal K}_{tree}(OA_b) & = & 2^4 \, (o_{16}+v_{16})\, ,\\
\label{TUOA} {\cal A}^{un}_{tree}(OA_b)&=& 2^{-6} [ (n_o + n_s)^2 \, o_{16}  +
(n_o- n_s)^2 \, v_{16} ]\, .
\end{eqnarray}  To obtain the  M\"obius amplitudes ${\cal M}_{tree}$ from
$K_{tree}$ and $A^{un}_{tree}$, one requires that each term in the power series
expansion of the total tree channel amplitude ${\cal K}_{tree}+{\cal A}_{tree}+
{\cal M}_{tree}$  be a perfect square. One gets 
\begin{eqnarray}
\label{MOB} {\cal M}_{tree}(OB_b) & = & \epsilon_1 \, (n_o +n_v + n_s +n_c)
\, {\hat o}_{16}\, ,\\
\label{M2B} {\cal M}_{tree}(IIB_b) & = & \epsilon_2 \, (n_o + n_v) \, {\hat
o}_{16} +
\epsilon_3 \, (n_o -n_v) \, {\hat s}_{16}\, , \\ \label{MOA} {\cal M}_{tree}(OA_b)
& = &
\epsilon_4 \, (n_o + n_s) \, {\hat o}_{16}  + \epsilon_5 \, (n_o- n_s) \, {\hat
v}_{16}\, , 
\end{eqnarray}  where $\epsilon_i = \pm 1$ will be determined by tadpole
conditions. The `hat' notation in the amplitudes Eqs.(\ref{MOB})-(\ref{MOA})
means that the overall phase present in the characters ${\hat {r}}_{16}$ is
dropped. This phase arises because the  modulus   over which ${\cal M}$ is
integrated (and which is not displayed here) is not purely imaginary but  is
shifted by 1/2,  inducing  in  the partition functions $i_{16}$  an alternate shift
of sign in its
 power series expansion as well as a global phase. This one half shift is needed 
 to preserve the group invariance of the amplitudes \cite{eht}. A  detailed
discussion of the shift in general cases  can be found
  in reference \cite{sagr}.

We now impose the tadpole conditions on the three theories, namely we
impose the cancellation of the divergences due to the massless mode
exchanges in the total amplitudes
${\cal K}_{tree}+{\cal A}_{tree}+ {\cal M}_{tree}$. 

We briefly review the results for $OB_b$ and $IIB_b$ theories already given in
detail in \cite{eht}. For $OB_b\,$, inspecting Eqs.(\ref{TKOB}), (\ref{TUOB}) and
(\ref{MOB}), we see that there are massless modes coming from the Dedekind
functions (describing the non-compact space) and
$o_{16}$ at level one,  and from
$s_{16}$ and $c_{16}$ at level zero. Taking into account the alternating signs
in the M\"obius amplitudes,   tadpole cancellation at level one implies
$\epsilon_1=1$  and 
$n_o+n_v+n_s+n_c=64$. Cancellation of the tadpole arising at  level zero
implies $n_o =n_v$ and $n_s =n_c$. Consequently the tadpole-free descendant
of $OB_b$ is characterized by a Chan-Paton group given by
$[SO(n) \times SO(32-n)]^2$. For $IIB_b\,$,  Eqs.(\ref{TK2B}), (\ref{TU2B}) and
(\ref{M2B}) yield tadpoles 
 from   the Dedekind functions and $o_{16}$ at level one,  and from
 $s_{16}$ at level 0. The first tadpole condition gives $\epsilon_2 =+1$ and
$n_o+n_v=32$, the second  gives $\epsilon_3 =-1$ and
$n_o-n_v=32$. Thus $n_v=0$ and the open descendant of
$IIB_b$ has the Chan-Paton group $SO(32)$. After truncation, the bosonic open
descendant of $IIB_b$ gives the anomaly-free Type~I theory and the bosonic
descendant of $OB_b$ gives the fermionic tadpole-free $[SO(n)
\times SO(32-n)]^2$ theory
\cite{eht}. 

We now turn to the construction of the descendant of $OA_b$ which was not
discussed in
\cite{eht}. The  Eqs.(\ref{TKOA}), (\ref{TUOA}) and (\ref{MOA}) yield only
tadpoles  from the Dedekind functions and
$o_{16}$  at level 1. The tadpole condition gives then $\epsilon_4 =+1$ and
$n_o + n_s =32$. There is no condition on $\epsilon_5$. The Chan-Paton group
is determined, as usual, by the massless vector contribution to the open
string partition function
${\cal A}+{\cal M}$. There are $(1/2)(32^2-32)$ such vectors. Consequently the
tadpole-free open descendant of $OA_b$ has the  Chan-Paton group
$SO(n) \times SO(32-n)$. After truncation of this theory one recovers the
open fermionic string theory with gauge group
$SO(n)
\times SO(32-n)$ which is the tadpole-free open descendant of the
10-dimensional type $OA$ theory discussed in references \cite{sagn1,bega}.

Starting with the bosonic tadpole-free open string theories and performing
the universal truncation, we thus recover all the  tadpole-free and
anomaly-free open fermionic string theories.

\section*{Acknowledgments}

This work was supported in part  by the NATO grant PST.CLG.979008.
A.C. thanks the Thai government for a DPST scholarship and A.T. 
thanks the Service de Physique Th\'eorique and  the Service de Physique
Th\'eorique  et Math\'ematique of the Universit\'e Libre de Bruxelles for their
warm hospitality.\vskip .5cm

\noindent 
We are grateful to Augusto Sagnotti for illuminating discussions.

\newpage

\appendix
\setcounter{equation}{0}
\renewcommand\theequation{\thesection.\arabic{equation}} 
\section{T-duality and  the Lorentz algebra} To obtain a D8-brane
from a  fermionic D9-brane in the light cone gauge, or equivalently in the
fermionic subspace of a bosonic string, one may compactify a non-compact
direction, say the 8th,   perform a T-duality and then decompactify. 
Compactification breaks the
$SO_{trans}(8)$ to $SO_{trans}(7)\times U(1)$ and  the $SO(8)$ generators
$L^{ab}\  (a,b = 1,2 ... 8)$ decompose in  $SO(7)$ generators $L^{ij}\  (i,j = 1,2 ...
7)$ and a vector $L^{i\, 8}$. One may then define separately the rotations 
$L_L^{i\, 8}$ and $ L_R^{i\, 8}$ in the left and right closed string sectors.
T-duality amounts to perform a parity operation in the left sector and hence
sends $L_L^{i\, 8}\to - L_L^{i\, 8}$,
$L_R^{i\,8}\to  L_R^{i\, 8}$. The light-cone
$so\,  (8)$ algebra of the truncated theory was identified with the diagonal
algebra $so_{diag}(8) = diag [so_{trans}(8)
\times so_{int}(8)]$.  The generators $J^{ab}$  of this algebra are given by
 \begin{equation}
\label{physgene} J^{ab}=L^{ab}+K_o^{ab}
\end{equation}
 where the operators $K_o^{ab}$ belong to  the internal
$so_{int}(8)$ and are the zero-modes of the full $\widehat{so}_{int}(8)$ affine
Lie algebra. Compactifying the 8th direction and performing  a T-duality does
not affect the right generators $J_R^{i8}$ of
$so_{diag}(8)$ associated with the 8th direction but implies that
 the left generators become
$J_L^{i8}=-L_L^{i8}+ K_{oL}^{i8}$. Inspecting the extension of
$so_{diag}(8)$ to $so_{diag}(9,1)$ described in
\cite{ens} one can see that the left algebra no longer closes.  The remedy is to
perform on the generator $K_{oL}^{ij}$ of the Kac-Moody algebra, {\it together
with } the T-duality,  the transformation 
\begin{eqnarray}
\label{tKgene} K_{oL}^{ij} &\rightarrow & K_{oL}^{ij} \, \, i,j=1..7\nonumber
\\ K_{oL}^{i\, 8} &\rightarrow & -K_{oL}^{i\, 8}\, . 
\end{eqnarray} The transformations Eq.(\ref{tKgene}) on the left
$\widehat{so}_{int}(8)$ Kac-Moody generators implies  that, when expressed
in the Cartan-Weyl basis one has for the fourth Cartan generator
${H_4}_L$
\footnote{Here the four Cartan generators
$H_i, \,i=1..4$ correspond to the 4 commuting rotations $K^{2i-1,2i}$.}
\begin{equation}
\label{cartang} {H_4}_L \rightarrow -{H_4}_L.
\end{equation} This implies a flip in sign of the fourth component of all the
vectors of the left lattice as compared to the right lattice. This maps the
$(o)_8$ and the $(v)_8$
 left lattices onto themselves but interchanges the $(s)_8$ and
$(c)_8$ lattices. Thus, as expected, to ensure closure of the left Lorentz
algebra in the fermionic subspace, one must accompany
the T-duality in the 8th dimension  by the switch
\begin{eqnarray}
\label{switch} OB &\Leftrightarrow & OA\, , \nonumber\\ IIB &\Leftrightarrow
& IIA\, .
\end{eqnarray}
Hence, to ensure Lorentz invariance in the truncation, a corresponding switch
must be made in  the bosonic parents  when a T-duality is performed  in the
bosonic theory. 
 
\newpage

\section{Lagrangian realizations of the theories}
\setcounter{equation}{0} We shall give explicit odd Lagrangian realizations of
the four bosonic theories which are related for $OB_b$,   $OA_b$ and $IIB_b$,
$IIA_b$ by odd E-dualities. 

We first list  basis vectors of a unit cell $\{2\, {\bf e_{ai}}\}$ 
 and of the lattice-dual cell $\{{\bf e^a_{\ i}}/2\}$ for the  
$OB_b$ and $OA_b$ theories and quote their
 volume.

\begin{center}
$\begin{array}{|c|c|}
\hline OB_b& OA_b\\
\hline \hline
\multicolumn{2}{|c|}{\{2{\bf \, e_{ai}}\} }\\
\hline 
\left[\begin{array}{cccccccc}
1&0&0&0&0&0&0&1\\0&1&0&0&0&0&0&1\\0&0&1&0&0&0&0&1\\
0&0&0&1&0&0&0&1\\0&0&0&0&1&0&0&1\\0&0&0&0&0&1&0&1\\
0&0&0&0&0&0&1&1\\0&0&0&0&0&0&0&2\\
\end{array}\right]&\left[\begin{array}{cccccccc}
1&\hbox{-}1&0&0&0&0&0&0\\0&1&\hbox{-}1&0&0&0&0&0\\0&0&1
&\hbox{-}1&0&0&0&0\\
0&0&0&1&\hbox{-}1&0&0&0\\0&0&0&0&1&\hbox{-}1&0&0\\0&0
&0&0&0&1&\hbox{-}1&0\\
0&0&0&0&0&0&1&\hbox{-}1\\0&0&0&0&0&0&0&1\\ 
\end{array}\right]\\

\hbox{Det} \{2{\bf \, e_{ai}}\}=2&\hbox{Det} \{2{\bf \, e_{ai}}\}=1\\
\hline
\hline
\multicolumn{2}{|c|}{\{{\bf \, e^a_{\ i}}/2\} }\\
\hline 
\left[\begin{array}{cccccccc}
1&0&0&0&0&0&0&0\\0&1&0&0&0&0&0&0\\0&0&1&0&0&0&0&0\\
0&0&0&1&0&0&0&0\\0&0&0&0&1&0&0&0\\0&0&0&0&0&1&0&0\\
0&0&0&0&0&0&1&0\\{\hbox{-}1\over2}&{\hbox{-}1\over2}&{\hbox{-}1\over2}&{\hbox{-}1\over2}&{\hbox{-}1\over2}
&{\hbox{-}1\over2}&{\hbox{-}1\over2}&{1\over2}\\
\end{array}\right]&\left[\begin{array}{cccccccc}
1&0&0&0&0&0&0&0\\1&1&0&0&0&0&0&0\\1&1&1&0&0&0&0&0\\
1&1&1&1&0&0&0&0\\1&1&1&1&1&0&0&0\\1&1&1&1&1&1&0&0\\
1&1&1&1&1&1&1&0\\1&1&1&1&1&1&1&1\\
\end{array}\right]\\

\hbox{Det} \{{\bf \, e^a_{\ i}}/2\}=1/2&\hbox{Det} \{{\bf \, e^a_{\ i}}/2\}=1\\
\hline
\end{array}$
\vskip.4cm

 Table V
\end{center}
\noindent  The dual cells $\{{\bf e^a_{\ i}}/2\}$ of  $OB_b$ and $OA_b$ are
respectively unit cells  of the weight lattice $(o)+(v)+(s)+(c)$ and  the
 lattice $(o)+(v)$ of
$SO(16)$, in accordance with the D9-brane tree amplitudes Eqs.(\ref{ATB})
given by ${\bf p_L} ={\bf p_R} = (1/2) m_a {\bf e^a}$.  The spectra of
D25-branes and  of closed strings depend on
$b_{ab}$ and we now write the metric tensors $g_{ab} = {\bf
\, e_a}.{\bf \, e_b}$ and the canonical $b_{ab}$ Eq.(\ref{oldb})  for both theories. 

\begin{center}

$\begin{array}{|c|c|}
\hline OB_b& OA_b\\
\hline \hline
\multicolumn{2}{|c|}{\{4g_{ab}\} }\\
\hline 
\left[\begin{array}{cccccccc}
2&1&1&1&1&1&1&2\\1&2&1&1&1&1&1&2\\1&1&2&1&1&1&1&2\\
1&1&1&2&1&1&1&2\\1&1&1&1&2&1&1&2\\1&1&1&1&1&2&1&2\\
1&1&1&1&1&1&2&2\\2&2&2&2&2&2&2&4\\
\end{array}\right]&\left[\begin{array}{cccccccc}
2&\hbox{-}1&0&0&0&0&0&0\\\hbox{-}1&2&\hbox{-}1&0&0&0&0&0\\0&\hbox{-}1&2&\hbox{-}1&0&0&0&0\\
0&0&\hbox{-}1&2&\hbox{-}1&0&0&0\\0&0&0&\hbox{-}1&2&\hbox{-}1&0&0\\0&0&0&0&\hbox{-}1&2&\hbox{-}1&0\\
0&0&0&0&0&\hbox{-}1&2&\hbox{-}1\\0&0&0&0&0&0&\hbox{-}1&1\\
\end{array}\right]\\
\hline
\hline
\multicolumn{2}{|c|}{{\{4b_{ab}\} } }\\
\hline 
\left[\begin{array}{cccccccc}
0&\hbox{\hbox{-}1}&\hbox{-}1&\hbox{\hbox{-}}1&\hbox{\hbox{-}}1&\hbox{\hbox{-}}1&\hbox{-}1&\hbox{-}2\\1&0&\hbox{-}1&\hbox{-}1&\hbox{-}1&\hbox{-}1&\hbox{-}1&\hbox{-}2\\
1&1&0&\hbox{-}1&\hbox{-}1&\hbox{-}1&\hbox{-}1&\hbox{-}2\\
1&1&1&0&\hbox{-}1&\hbox{-}1&\hbox{-}1&\hbox{-}2\\1&1&1&1&0&\hbox{-}1&\hbox{-}1&\hbox{-}2\\1&1&1&1&1&0&\hbox{-}1&\hbox{-}2\\
1&1&1&1&1&1&0&\hbox{-}2\\2&2&2&2&2&2&2&0     \\
\end{array}\right]&\left[\begin{array}{cccccccc}
0&1&0&0&0&0&0&0\\\hbox{-}1&0&1&0&0&0&0&0\\
0&\hbox{-}1&0&1&0&0&0&0\\
0&0&\hbox{-}1&0&1&0&0&0\\0&0&0&\hbox{-}1&0&1&0&0\\0&0&0&0&\hbox{-}1&0&1&0\\
0&0&0&0&0&\hbox{-}1&0&1\\0&0&0&0&0&0&\hbox{-}1&0     \\
\end{array}\right]\\
\hline
\end{array}$

\vskip .5 cm
 Table VI
\end{center}

\noindent We can now write the E-dual basis $\{{\bf E^a_{\ i}}\} $ of $\{{\bf
e_{ai}}\}$ and its lattice-dual $\{{\bf E_{ai}}\}$ given by Eqs.(\ref{edef}) and
(\ref{dualbasis}). The basis chosen for the unit $OB_b$ cell contains a 
 root lattice vector of squared length 4. As expected from   the discussion in
Section 5,  the realizations of  $OB_b$ and 
$OA_b$ of Table V are both odd. We indeed see  by comparing  Table V  with
Table VII  that, in the particular choice of unit cells taken here,   $\{{\bf
e_{ai}}\}_{OB_b}$ =
$\{{\bf E^a_{\ i}}\}_{OA_b}$ and
$\{{\bf e_{ai}}\}_{OA_b}$ = $\{{\bf E^a_{\ i}}\}_{OB_b}$. It is easily verified from
these tables  by rewriting Eq.(\ref{close}) as
\begin{eqnarray} {\bf p_R} &=& {1\over2} (m_a {\bf e^a} + n^a{\bf E_a} )\, , 
\nonumber \\ {\bf p_L} &=& {\bf p_R} - 2n^a {\bf e_a}\ , \label{eclosed}
\end{eqnarray}   that we indeed correctly get the closed string spectra of
both theories.
 Finally,  the tree channel amplitude of the D25-brane of each  theory  (given
by $m_a=0$) is nothing else than the D9 tree channel amplitude  of the other
theory, as it should  be in odd realizations.
\vskip .5cm
\begin{center}
$\begin{array}{|c|c|}
\hline OB_b& OA_b\\
\hline \hline
\multicolumn{2}{|c|}{\{{\bf \, E_{ai}}/2\} }\\
\hline 
\left[\begin{array}{cccccccc}
1&0&0&0&0&0&0&0\\1&1&0&0&0&0&0&0\\1&1&1&0&0&0&0&0\\
1&1&1&1&0&0&0&0\\1&1&1&1&1&0&0&0\\1&1&1&1&1&1&0&0\\
1&1&1&1&1&1&1&0\\1&1&1&1&1&1&1&1\\
\end{array}\right]&\left[\begin{array}{cccccccc}
1&0&0&0&0&0&0&0\\0&1&0&0&0&0&0&0\\0&0&1&0&0&0&0&0\\
0&0&0&1&0&0&0&0\\0&0&0&0&1&0&0&0\\0&0&0&0&0&1&0&0\\
0&0&0&0&0&0&1&0\\{\hbox{-}1\over2}&{\hbox{-}1\over2}&{\hbox{-}1\over2}&{\hbox{-}1\over2}&{\hbox{-}1\over2}
&{\hbox{-}1\over2}&{\hbox{-}1\over2}&{1\over2}\\
\end{array}\right]\\

\hbox{Det} \{{\bf \, E_{ai}}/2\}=1&\hbox{Det} \{{\bf \, E_{ai}}/2\}=1/2\\
\hline
\hline
\multicolumn{2}{|c|}{\{2{\bf \, E^a_{\ i}}\} }\\
\hline 
\left[\begin{array}{cccccccc}
1&\hbox{-}1&0&0&0&0&0&0\\0&1&\hbox{-}1&0&0&0&0&0\\0&0&1
&\hbox{-}1&0&0&0&0\\
0&0&0&1&\hbox{-}1&0&0&0\\0&0&0&0&1&\hbox{-}1&0&0\\0&0
&0&0&0&1&\hbox{-}1&0\\
0&0&0&0&0&0&1&\hbox{-}1\\0&0&0&0&0&0&0&1\\ 
\end{array}\right] &\left[\begin{array}{cccccccc}
1&0&0&0&0&0&0&1\\0&1&0&0&0&0&0&1\\0&0&1&0&0&0&0&1\\
0&0&0&1&0&0&0&1\\0&0&0&0&1&0&0&1\\0&0&0&0&0&1&0&1\\
0&0&0&0&0&0&1&1\\0&0&0&0&0&0&0&2\\
\end{array}\right]\\

\hbox{Det} \{2{\bf \, E^a_{\ i}}\}=1&\hbox{Det} \{2{\bf \, E^a_{\ i}}\}=2\\
\hline
\end{array}$
\vskip.4cm

 Table VII
\end{center}

 We now present in Table VIII the   basis vectors of a unit cell $\{{\bf e_{ai}}\}$ 
and of the lattice-dual cell
$\{{\bf e^{ai}}\}$ for the  $IIB_b$ and $IIA_b$ theories. Their metric tensor
$g_{ab}$ is also displayed but the antisymmetric tensor $b_{ab}$,   which  has
the canonical form,  is omitted. Again we have chosen a unit cell of
$IIB_b$ containing a vector of length squared 4 and we have odd realizations.
The E-dual bases are shown in Table IX. As in the previous
$OB_b-OA_b$ case we get, comparing  Table~VIII  with Table IX,  $\{{\bf
e_{ai}}\}_{IIB_b}$ =
$\{{\bf E^a_{\ i}}\}_{IIA_b}$ and
$\{{\bf e_{ai}}\}_{IIA_b}$ = $\{{\bf E^a_{\ i}}\}_{IIB_b}$. One may check as
previously that the closed string spectrum, as well as  the D9 and D25-brane
spectra characteristic of odd realizations,  are correctly given.

\begin{center}
$\begin{array}{|c|c|}
\hline IIB_b& IIA_b\\
\hline \hline
\multicolumn{2}{|c|}{\{2{\bf \, e_{ai}}\} }\\
\hline 
\left[\begin{array}{cccccccc} 1&0&0&0&0&0&\hbox{-}1&0\\
0&1&0&0&0&0&\hbox{-}1&0\\ {\hbox{-}1\over2}&{\hbox{-}1\over2}
&{1\over2}
&{\hbox{-}1\over2}&{\hbox{-}1\over2}&{\hbox{-}1\over2}&{1\over2}&{\hbox{-}1\over2}\\
0&0&0&1&0&0&\hbox{-}1&0\\{\hbox{-}1\over2}&{\hbox{-}1\over2}&{\hbox{-}1\over2}&{\hbox{-}1\over2}
&{1\over2}&{\hbox{-}1\over2}&{1\over2}&{\hbox{-}1\over2}\\0&0&0&0&0&1&\hbox{-}1&0\\
{\hbox{-}1\over2}&{\hbox{-}1\over2}&{\hbox{-}1\over2}&{\hbox{-}1\over2}&{\hbox{-}1\over2}&{\hbox{-}1\over2}
&{3\over2}&{\hbox{-}1\over2}\\
{1\over2}&{1\over2}&{1\over2}&{1\over2}&{1\over2}
&{1\over2}&{1\over2}&{1\over2}\\  
\end{array} \right]&\left[\begin{array}{cccccccc}
1&\hbox{\hbox{-}}1&0&0&0&0&0&0\\0&1&1&0&0&0&0&0\\0&0&1
&1&0&0&0&0\\ 0&0&0&1&1&0&0&0\\0&0&0&0&1&1&0&0\\0&0
&0&0&0&1&1&0\\
0&0&0&0&0&0&1&0\\{1\over2}&{1\over2}&{1\over2}&{1\over2}&{1\over2}
&{1\over2}&{1\over2}&{1\over2}\\  
\end{array}  \right]\\
\hbox{Det} \{2{\bf \, e_{ai}}\}=1&\hbox{Det} \{2{\bf \, e_{ai}}\}=1/2\\
\hline
\hline
\multicolumn{2}{|c|}{\{{\bf \, e^a_{\ i}}/2\} }\\
\hline 
\left[ \begin{array}{cccccccc}
1&0&0&0&0&0&0&\hbox{-}1\\0&1&0&0&0&0&0&\hbox{-}1\\0&0&1&0&0&0&0&\hbox{-}1\\
0&0&0&1&0&0&0&\hbox{-}1\\0&0&0&0&1&0&0&\hbox{-}1
\\0&0&0&0&0&1&0&\hbox{-}1\\
{1\over2}&{1\over2}&{\hbox{-}1\over2}&{1\over2}&{\hbox{-}1\over2}&{1\over2}&
{1\over2}&{\hbox{-}3\over2}\\{1\over 2}&{1\over2}&{1
\over2}&{1\over2}&{1\over2} &{1\over2}&{1\over2}&{\hbox{-}3\over2}\\  
\end{array}\right]&\left[\begin{array}{cccccccc}
1&0&0&0&0&0&0&\hbox{-}1\\1&1&0&0&0&0&0&\hbox{-}2\\\hbox{-}1&\hbox{-}1&1&0&0&0&0&1\\
1&1&\hbox{-}1&1&0&0&
0&\hbox{-}2\\\hbox{-}1&\hbox{-}1&1&\hbox{-}1&1&0&0&1\\1&1&\hbox{-}1&1&\hbox{-}1&1&0&\hbox{-}2\\
\hbox{-}1&\hbox{-}1&1&\hbox{-}1&1&\hbox{-}1&1&1
\\0&0&0&0&0&0&0&2\\ 
\end{array}\right]\\ 
\hbox{ Det} \{{\bf \, e^a_{\ i}}/2\}=1&\hbox{Det} \{{\bf \, e^a_{\ i}}/2\}=2\\
\hline
\hline
\multicolumn{2}{|c|}{\{4g_{ab}\} }\\
\hline 
\left[\begin{array}{cccccccc}
2&1&\hbox{-}1&1&\hbox{-}1&1&\hbox{-}2&0\\1&2&\hbox{-}1&1&\hbox{-}1&1&\hbox{-}2&0\\\hbox{-}1&\hbox{-}1&2&\hbox{-}1&1&\hbox{-}1
&2&\hbox{-}1\\
1&1&\hbox{-}1&2&\hbox{-}1&1&\hbox{-}2&0\\\hbox{-}1&\hbox{-}1&1&\hbox{-}1&2&\hbox{-}1&2&\hbox{-}1\\
1&1&\hbox{-}1&1&\hbox{-}1&2&\hbox{-}2&0\\
\hbox{-}2&\hbox{-}2&2&\hbox{-}2&2&\hbox{-}2&4&\hbox{-}1\\0&0&\hbox{-}1&0&\hbox{-}1&0&\hbox{-}1&2\\
\end{array}\right]&\left[\begin{array}{cccccccc}
2&\hbox{-}1&0&0&0&0&0&0\\\hbox{-}1&2&1&0&0&0&0&1
\\0&1&2&1&0&0&0&1\\
0&0&1&2&1&0&0&1\\0&0&0&1&2&1&0&1\\0&0&0&0&1&2&1&1\\
0&0&0&0&0&1&1&{1\over2}\\0&1&1&1&1&1&{1\over2}&2\\
\end{array}\right]\\
\hline
\end{array}$

\vskip .5 cm
 Table VIII
\end{center}

\begin{center}
$\begin{array}{|c|c|}
\hline IIB_b& IIA_b\\
\hline \hline
\multicolumn{2}{|c|}{\{{\bf \, E_{ai}}/2\} }\\
\hline 
\left[\begin{array}{cccccccc}
1&0&0&0&0&0&0&\hbox{-}1\\1&1&0&0&0&0&0&\hbox{-}2\\\hbox{-}1&\hbox{-}1&1&0&0&0&0&1\\
1&1&\hbox{-}1&1&0&0&
0&\hbox{-}2\\\hbox{-}1&\hbox{-}1&1&\hbox{-}1&1&0&0&1\\1&1&\hbox{-}1&1&\hbox{-}1&1&0&\hbox{-}2\\
\hbox{-}1&\hbox{-}1&1&\hbox{-}1&1&\hbox{-}1&1&1
\\0&0&0&0&0&0&0&2\\
\end{array}\right]&\left[\begin{array}{cccccccc}
1&0&0&0&0&0&0&\hbox{-}1\\0&1&0&0&0&0&0&\hbox{-}1\\0&0&1&0&0&0&0&\hbox{-}1\\
0&0&0&1&0&0&0&\hbox{-}1\\0&0&0&0&1&0&0&\hbox{-}1\\0&0&0&0&0&1&0&\hbox{-}1\\
{1\over2}&{1\over2}&{\hbox{-}1\over2}&{1\over2}&{\hbox{-}1\over2}&{1\over2}&
{1\over2}&{\hbox{-}3\over2}\\{1\over2}&{1\over2}&{1
\over2}&{1\over2}&{1\over2} &{1\over2}&{1\over2}&{\hbox{-}3\over2}\\ 
\end{array}\right]\\
\hbox{Det} \{{\bf \, E_{ai}}/2\}=2&\hbox{Det} \{{\bf \, E_{ai}}/2\}=1\\
\hline
\hline
\multicolumn{2}{|c|}{\{2{\bf \, E^a_{\ i}}\} }\\
\hline 
\left[\begin{array}{cccccccc}
1&\hbox{\hbox{-}}1&0&0&0&0&0&0\\0&1&1&0&0&0&0&0\\0&0&1
&1&0&0&0&0\\ 0&0&0&1&1&0&0&0\\0&0&0&0&1&1&0&0\\0&0
&0&0&0&1&1&0\\
0&0&0&0&0&0&1&0\\{1\over2}&{1\over2}&{1\over2}&{1\over2}&{1\over2}
&{1\over2}&{1\over2}&{1\over2}\\  
\end{array}  \right] &\left[\begin{array}{cccccccc}
1&0&0&0&0&0&\hbox{-}1&0\\ 0&1&0&0&0&0&\hbox{-}1&0\\
{\hbox{-}1\over2}&{\hbox{-}1\over2} &{1\over2}
&{\hbox{-}1\over2}&{\hbox{-}1\over2}&{\hbox{-}1\over2}&{1\over2}&{\hbox{-}1\over2}\\
0&0&0&1&0&0&\hbox{-}1&0\\{\hbox{-}1\over2}&{\hbox{-}1\over2}&{\hbox{-}1\over2}&{\hbox{-}1\over2}
&{1\over2}&{\hbox{-}1\over2}&{1\over2}&{\hbox{-}1\over2}\\0&0&0&0&0&1&\hbox{-}1&0\\
{\hbox{-}1\over2}&{\hbox{-}1\over2}&{\hbox{-}1\over2}&{\hbox{-}1\over2}&{\hbox{-}1\over2}&{\hbox{-}1\over2}
&{3\over2}&{\hbox{-}1\over2}\\
{1\over2}&{1\over2}&{1\over2}&{1\over2}&{1\over2}
&{1\over2}&{1\over2}&{1\over2}\\  
\end{array} \right]\\
\hbox{Det} \{2{\bf \, E^a_{\ i}}\}=1/2&\hbox{Det} \{2{\bf \, E^a_{\ i}}\}=1\\
\hline
\end{array}$
\vskip.5cm

 Table IX
\end{center}
\newpage

 \section{Properties of E-duality}
\setcounter{equation}{0}
 Here we will prove that E-duality  {\it always} maps
\begin{eqnarray}
\label{echange} OA_b & \rightarrow & OB_b \, ,\nonumber\\ IIA_b &
\rightarrow & IIB_b\, .
\end{eqnarray}

It suffices to show that  one can not have
$OA_b
\rightarrow OA_b$ and $ IIA_b \rightarrow IIA_b\,$, as the only different
theory whose spectrum  matches the   $OA_b$ ($ IIA_b$)  theory is the
$OB_b$ ($ IIB_b$) theory.  To prove these statements we  use  an identity
between 
$\{{\bf e_a}\}$ defined in Eqs.(\ref{ptorus})-(\ref{pmetric}) and  
$\{{\bf E_a}\}$ defined in Eq.(\ref{dualbasis}). One has \begin{equation}
\label{ident} {\bf E_a. e_b}+ {\bf E_b. e_a}= 4\,  {\bf e_a. e_b}\, .
\end{equation} It is convenient to work with the vectors ${\bf W_a}=(1/2)\, 
{\bf E_a}$ and ${\bf w_a}= 2\, {\bf e_a}$  which belong to some sublattices $(r)$
of the $SO(16)$ weight lattice
 (see Eqs.(\ref{pbase}) and (\ref{edualbis})\,). One has
\begin{equation}
\label{identbis} {\bf W_a. w_b}+ {\bf W_b.w_a}=  {\bf w_a.w_b}\, .
\end{equation}
\vskip.2 cm
\underline{a) $OA_b \neq > OA_b$}

\vskip.2 cm
\noindent
If one assumes $OA_b \rightarrow OA_b$ then,     using Eqs.(\ref{pbase})-
(\ref{sublattice}),  the sublattice is $(o)+(v)$ and  at least one $\bf w_a$
belongs to $(v)$. For this $\bf w_a\,$, Eq.(\ref{identbis}) becomes, picking
$a=b$,
\begin{equation}
\label{identer} {\bf W_a. w_a}+ {\bf W_a. w_a}=  {\bf w_a.w_a}.
\end{equation} But, were $OA_b$ the E-dual of $OA_b$,   ${\bf W_a}$  would 
also belong to $(o)+(v)$.  This is impossible because the l.h.s of
Eq.(\ref{identer}) would be even and the r.h.s would be odd.

\vskip.2 cm
\underline{b) $IIA_b \neq > IIA_b$}
\vskip.2 cm
\noindent
If $IIA_b$ would be mapped onto itself,  the ${\bf W_a}$ would belong to the
root lattice
$(o)$ as they are the lattice-dual of ${\bf W^a}$ which would form a basis of the 
$SO(16)$ weight lattice. Introducing 
 fundamental weights ${\bf v_a}$ dual to the simple roots ${\bf r^b}$
 one  writes
\begin{eqnarray} {\bf W_a} & = & n_{ac}{\bf r^c}\, , \nonumber \\ {\bf w_b} & =
& m_b^c{\bf v_c}\, , \end{eqnarray} and hence
\begin{equation}
\label{interm} {\bf W_a .w_b}= n_{ac}m_b^c \in {\cal Z} ~~~ 
\forall a,b \, . \end{equation} Using Eqs.(\ref{identbis}) and (\ref{interm}) we
would then find that
\begin{equation}
\label{cond} {\bf w_a. w_b} \in {\cal Z}~~~\forall a,b\, .
\end{equation} But the vectors
$ {\bf w_a}$ form a basis of $IIA_b$ which must at least contain a couple of
weight vectors $({\bf w_a,w_b})= (v,s)$ or
$(v,c)$ or
$(c,s)$. For such couples, the scalar product in  Eq.(\ref{cond}) is  half-integer.

\end{document}